# Statistical Analysis of Droplet Size Distributions in Liquid-Jet-in-Crossflow Atomization

Tom Johny[1], Bharat Bhatia[2], Zafar Alam[1], Ashoke De[1]*

[1] *Department of Aerospace Engineering, Indian Institute of Technology Kanpur, 208016, Kanpur, India.*

[2] *Current affiliation: Department of Mechanical Engineering, Eindhoven University of Technology, Eindhoven 5600MB, Netherlands.*

This study investigates the droplet size distribution (DSD) and atomization characteristics of a liquid jet injected into a crossflow (LJICF) under varied momentum flux ratios, Weber numbers, and flow conditions. Numerical simulations are performed using the validated compressible Volume of Fluid-Lagrangian Particle Tracking (VOF-LPT) coupled framework to capture both the primary and secondary atomization processes. Key parameters, including momentum flux ratio, Weber number, crossflow pressure, and velocity, were analyzed to assess their impact on droplet size characteristics, including Sauter mean diameter (SMD) and standard deviation (STD) in the downstream region. The discrete size distribution of droplets comprising probability density and cumulative distribution reveals a shift toward finer, more uniform droplets under enhanced breakup conditions. The findings emphasize the critical role of aerodynamic forces and instabilities in driving efficient atomization, with higher momentum flux ratios and Weber numbers leading to finer and more uniform droplets. Increased crossflow pressure promotes finer droplet formation but is found to reduce droplet density in the downstream domain due to a confined spray plume, delayed particle conversion, and reduced droplet residence time. The log-normal and Rosin-Rammler distributions effectively capture droplet size trends, with the former closely representing the skewness and tail behavior and the latter accurately representing intermediate and larger droplets. However, both have limitations in replicating sharp peaks and the smallest droplet sizes, respectively.

**NOMENCLATURE**

| | |
|---|---|
| $c_p$ | Specific heat of the gas (J/Kg.K) |
| $D_{32}$ | Sauter mean diameter (μm) |
| $D_{10}$ | Arithmetic mean diameter (μm) |
| $D_N$ | Nozzle diameter |
| $V_j$ | Liquid jet velocity (m/s) |
| $V_a$ | Crossflow velocity (m/s) |
| We | Weber number |
| $\mu_{air}$ | Dynamic viscosity of air (N.s/m$^2$) |
| $\mu_{jet}$ | Dynamic viscosity of liquid (N.s/m$^2$) |
| P | Pressure (Pa) |
| q | Momentum flux ratio |
| $Re_p$ | Reynolds number |
| $Sc$ | Schmidt number |
| $Sh$ | Sherwood number |
| c | Speed of sound (m/s) |

**Greek symbols**

| | |
|---|---|
| $\alpha_1$ | Volume fraction of liquid |
| $\sigma$ | Surface tension (N/m) |
| $\mu$ | Dynamic viscosity (N.s/m$^2$) |
| $\rho$ | Mixture density (Kg/m$^3$) |
| $k$ | Thermal conductivity of gas (W/m.K) |
| $\gamma$ | Specific heat ratio |

**Abbreviations**

| | |
|---|---|
| SMD | Sauter Mean Diameter (μm) |
| STD | Standard deviation |
| DSD | Droplet size distribution |
| PDP | Probability density plot |
| CDP | Cumulative distribution plot |
| LSA | Linear stability analysis |

[a] Electronic mail: ashoke@iitk.ac.in

## I. INTRODUCTION

Droplet statistical analysis is essential for understanding and characterizing sprays, particularly in scenarios where liquid jets interact with high-velocity, high-temperature crossflows at elevated pressures. This phenomenon is critical in various engineering applications, including fuel injection systems and advanced propulsion technologies such as gas turbines, afterburners, augmenters, and ramjet–scramjet combustors. The evaporation and subsequent combustion of fuel droplets in these systems are strongly influenced by the droplet size distribution resulting from the breakup and atomization process. Errors in predicting droplet size distribution (DSD) can have significant engineering consequences. For instance, in scramjet combustors, inaccurate DSD prediction may lead to non-uniform evaporation and incomplete mixing[66], which can trigger combustion instabilities and local flame blowout, or excessive pressure oscillations, all of which reduce engine efficiency and elevate pollutant emissions. In gas turbines, underestimating the prevalence of large droplets can cause fuel-rich zones, incomplete combustion, increased unburned hydrocarbon emissions, and hotspots that damage turbine components[65]. In afterburners and augmenters, erroneous DSD modeling can result in excessive smoke formation or flame quenching due to improper fuel-air mixing. Additionally, combustion instability in liquid-fueled systems has been directly linked to the interplay between droplet size, atomization, and dynamic flow oscillations, highlighting the need for accurate DSD modeling to maintain stable and clean combustion[67]. A comprehensive understanding of the statistical parameters governing droplet behavior is, therefore, crucial for the accurate analysis and optimization of such flows. Droplet size plays a significant role in influencing evaporation rates, mixing efficiency, and combustion performance[28]. Smaller droplets with higher surface area-to-volume ratios, evaporate more rapidly, enhancing efficient fuel-air mixing, which is critical for stable and complete combustion[11,26]. In applications such as spray cooling, droplet size distribution affects the cooling efficiency and uniformity, which are essential for effective thermal management[27]. Droplet size distribution is an important parameter in governing the flow of dispersed droplets in multiphase flows involving spray atomizations. Therefore, optimizing droplet size distribution is critical for improving spray performance and ensuring overall system efficiency and reliability across various spray applications.

The breakup dynamics of a liquid jet in crossflow (LJICF) is inherently complex, driven by interactions between the injected liquid jet and the perpendicular crossflow, leading to the primary and secondary atomization resulting in droplet formation[1,6,57]. The breakup process and resulting droplet size distribution are influenced by key parameters such as crossflow velocity, momentum flux ratio, and Weber number[22,29]. Studies have shown that increasing the crossflow Weber number, which quantifies the balance between inertial and surface tension forces, enhances liquid jet breakup, producing a broader range of droplet sizes[10,32,39].

Numerical simulations have become an essential tool for studying the complex breakup dynamics in liquid jets in crossflow (LJICF), providing detailed insights into droplet formation under controlled conditions[32,60]. The Volume of Fluid (VOF) method[24,50] is particularly effective at capturing the initial gas-liquid interface during primary jet breakup, while Lagrangian Particle Tracking (LPT) accurately resolves smaller droplets further downstream. The LPT framework, based on the PSI-CELL model[13], enables accurate gas-droplet flows by coupling particle tracking with fluid dynamics. Subsequent refinements extended LPT's applicability, including its adaptation for heat and mass transfer in combustion chambers[19], modeling of high-pressure fuel sprays[46] , and incorporation of particle-turbulence interactions through two-way coupling[14] . The VOF-LPT coupled framework integrates both methods, offering a comprehensive representation of droplet formation, atomization, and transport in LJICF[6,21,49,56,59,61]. This approach effectively captures both the primary jet breakup and the downstream evolution of individual droplets, making it well-suited for analyzing droplet dynamics in complex LJICF scenarios.

Experimental and computational approaches have significantly advanced our understanding of droplet dynamics in LJICF applications. Techniques such as Phase Doppler Interferometry (PDI) and high-speed imaging enable detailed measurements of droplet sizes, velocities, and trajectories. However, these methods often encounter limitations in resolution and sampling frequency, particularly in high-speed, dense spray regions[16,31,41]. To address these challenges, computational frameworks such as VOF-LPT are increasingly employed to complement experiments, providing a more comprehensive analysis of droplet behavior in LJICF. Computational models have been used to investigate droplet breakup mechanisms[5], while coupled frameworks capture detailed droplet size characteristics under varying crossflow conditions. Despite extensive research on droplet breakup, high-speed imaging, and computational modeling of spray formation and dynamics, limited studies explicitly integrate experimental data with computational insights on droplet statistics using VOF-LPT.

The primary and secondary atomization processes generate a large number of droplets with varying sizes, resulting in a non-uniform droplet size distribution (DSD)[20]. Characterizing these droplet size variations is essential for evaluating atomization quality and spray performance. Researchers used mathematical distribution functions to represent the DSD, which standardizes droplet size representation, smoothens out statistical fluctuations, and enables meaningful comparisons across different spray conditions. Commonly employed theoretical distributions include the Normal, Log-Normal, and Maximum Entropy distributions.[53] Empirical distribution functions derived from experimental data, such as the Nukiyama-Tanasawa,[37] Rosin-Rammler,[48] and the upper-limit function distributions[36], play a significant role in describing DSDs. Lefebvre and McDonell[28] particularly recommended the Rosin-Rammler distribution for its effectiveness in representing broad droplet size distributions in industrial spray applications. Heywood[23] further emphasized that selecting an appropriate distribution function provides valuable insights into droplet spread and uniformity, which are crucial for optimizing spray performance. Recent studies have introduced a fractal distribution model for characterizing atomized droplets in crossflow, where droplet size distribution is represented by a fractal dimension (FD) that adapts to spray pressure and flow conditions[15]. This approach offers a standalone metric for assessing droplet uniformity and size, extending beyond conventional parameters such as SMD and SPAN, with promising applications in air purification, agricultural spraying, and combustion.

Studies indicate that droplet size distributions in LJICF systems evolve downstream due to the influence of aerodynamic forces, coalescence, and secondary breakup. Prakash et al. [42] demonstrated that varied entry conditions impact dispersion, with turbulence promoting finer atomization. Recent studies have investigated the influence of injection orientation and orifice geometry on droplet size distributions and atomization characteristics of liquid jets in crossflow. Specifically, backward-injected jets and reduced injection angles intensify aerodynamic interactions, thereby promoting faster breakup and producing smaller droplet sizes due to enhanced shear and instability-driven mechanisms[63]. Additionally, variations in orifice geometry, such as elliptical and square discharge shapes, significantly affect jet trajectories, breakup locations, and droplet sizes, with elliptical orifices oriented horizontally resulting in finer droplets and narrower distributions compared to circular or square geometries[64]. Bai et al. [4] found that turbulent crossflows lead to wider dispersion and a broader droplet size spectrum, particularly in low-density flows. Shinjo and Umemura [55] observed that as droplets interact with the crossflow, the mean droplet size decreases while the distribution width increases. Further investigations by Bodoc et al.[8] and Prakash et al. [43] confirmed these downstream variations, especially under swirling or oscillating crossflows, which introduce additional heterogeneity in droplet sizes. This further highlights the role of swirl intensity and gas flow dynamics in enhancing droplet breakup and dispersion, adding complexity to the downstream droplet size distribution. High-temperature and high-pressure studies of JICF, like those by Amighi and Ashgriz[2,3] highlight the influence of environmental factors on atomization and overall droplet size characteristics. Such variations are critical in applications like combustion and cooling, where precise control over droplet dispersion and size is essential for performance optimization.

Despite significant advancements in understanding LJICF atomization, gaps remain in the literature regarding the comprehensive comparison between computational predictions and experimental data, particularly in fitting DSDs. Farvardin and Dolatabadi[17] highlighted the challenges in reconciling computational and experimental results for biodiesel jets, revealing discrepancies in droplet size predictions. Mashayek and Ashgriz [34] provided foundational insights into how factors like Weber number, crossflow velocity, and momentum flux, shape DSDs in LJICF. However, many studies rely on a limited set of distribution functions that may not fully capture the complex droplet dynamics observed in these scenarios. Furthermore, there is a limited understanding of the effects of momentum flux ratio, crossflow velocity, crossflow pressure, and Weber number on the cumulative, size, and volume distribution of spray droplets under these varied conditions. Addressing these gaps would enhance the accuracy of droplet size predictions in LJICF applications and improve the reliability of numerical models.

This study aims to provide a comprehensive investigation into the droplet size distribution (DSD) and atomization characteristics of a liquid jet in crossflow (LJICF) under varied flow conditions, including different momentum flux ratios, Weber numbers, crossflow pressures, and crossflow velocities. Using a validated VOF-LPT coupled framework, the study captures both primary and secondary atomization processes, enabling detailed analysis of droplet size characteristics such as SMD and STD. The work focuses on evaluating the influence of aerodynamic forces and instabilities on droplet breakup and dispersion, emphasizing the role of key flow parameters in shaping the droplet size spectrum. The study also examines probability density functions, cumulative distributions, and droplet velocity distributions to assess downstream variations in droplet characteristics under different flow conditions. Furthermore, by fitting droplet size data to established distribution functions, this research provides critical insights into the statistical representation of droplets and evaluates the effectiveness of these functions in characterizing the polydisperse nature of liquid sprays in cross-flow conditions.

## II. NUMERICAL METHODOLOGY

This study employs a validated compressible VOF-LPT coupled framework[6] to simulate Liquid Jet in Crossflow (LJICF). The VOF method captures liquid jet core deformation and primary breakup, while the LPT framework models secondary droplet breakup. Coupling is achieved through an image processing-Connected Component Labeling (CCL) algorithm adapted from Heinrich and Schwarze[21]. The iso-Advector approach[47] is used for interface reconstruction and advection. The following sections outline the governing equations for both Eulerian and Lagrangian frameworks.

### A. Eulerian Framework

The governing equations for mass, momentum, species, and energy are solved within an Eulerian framework to model the two-phase flow consisting of compressible, immiscible fluids. The VOF method tracks the liquid-gas interface, treating the mixture as a single fluid through a step function that represents the volume fraction of each phase within a computational cell:

$$\alpha_1 = \begin{cases} 0 & phase\ 2\ \text{(Gas)} \\ 0 < \alpha_1 < 1 & interface \\ 1 & phase\ 1\ \text{(Liquid)} \end{cases} \tag{1}$$

The mass conservation equation for phase $\boldsymbol{k}$ is:

$$\frac{\partial(\alpha_k \rho_k)}{\partial t} + \nabla.(\alpha_k \rho_k \mathbf{u}) = S_\rho\big|_L + S_\rho\big|_E \tag{2}$$

where $\mathbf{u}$ is the velocity, $S_\rho\big|_L$ accounts for mass transfer due to the vaporization of Lagrangian droplets[62], and $S_\rho\big|_E$ represents the mass transfer from the liquid to the gas phase. The integral form of the continuity equation is employed to calculate surface advection, while the total mass balance is expressed in differential form.

$$\frac{\partial(\rho)}{\partial t} + \nabla.(\rho \mathbf{u}) = S_\rho\big|_L \tag{3}$$

For cells involving liquid-gas calculations, fluid properties are computed as the weighted average of the phase fraction α for each computational cell.

$$\rho = \alpha_1 \rho_1 + \alpha_2 \rho_2 \tag{4a}$$

$$\mu = \alpha_1 \mu_1 + \alpha_2 \mu_2 \tag{4b}$$

where $\rho_1$, $\rho_2$ and $\mu_1$, $\mu_2$ are the density and dynamic viscosity for phases 1 and 2, respectively. The velocity field is solved using a single momentum transport equation:

$$\frac{\partial(\rho u)}{\partial t} + \nabla.(\rho \mathbf{u}\mathbf{u}) - \nabla \tau_{eff} = -\nabla pd - (\nabla \rho)\mathbf{g}.\mathbf{x} + F_{ST} + S_{\rho u}\big|_L + S_{\rho u}\big|_E \tag{5}$$

where $p_d$ is the piezometric pressure, $\tau_{eff}$ is the effective stress tensor, $\mu_{eff}$ the effective viscosity volume averaged as: $\mu_{eff} = \frac{4(\alpha_1\mu_1 + \alpha_2\mu_2)}{3}$, and $\rho$ in all the combined equations is the mixture density for each computational cell. $S_{\rho u}\big|_E, S_{\rho u}\big|_L$ are the Eulerian and Lagrangian source terms for momentum equations, which are generated because of the evaporation and the atomization of droplets in the domain. Using the Continuum Surface Force (CSF) model by Brackbill et al. [9] the surface tension force $F_{ST}$ is calculated as:

$$\vec{F}_{ST}(\vec{x}) = \sigma \int_S \kappa(\vec{y})\hat{n}(\vec{y})\delta(\vec{x}-\vec{y})dS \tag{6}$$

where σ is the liquid's surface tension, κ is the interface curvature, $\hat{n}$ is the unit normal, and δ is the Dirac-delta function at position vectors at $\vec{x}$ and $\vec{y}$. The curvature and normal are defined as $\kappa = -\nabla.\hat{n}$ , $\hat{n} = \frac{\vec{n}}{|\vec{n}|}$, where $\vec{n} = \nabla\alpha$. The surface tension source term is expressed as:

$$F_{ST} = \sigma \frac{\rho\kappa\nabla\alpha}{\frac{1}{2}(\rho_l+\rho_g)} \tag{7}$$

For the gas phase, the isentropic equation of state is used[35].

$$\frac{p}{\rho^\gamma} = a_c = \text{constant} \tag{8}$$

$$\left(\frac{\partial\rho}{\partial p}\right)_s = \frac{1}{a_c\gamma}\left(\frac{p}{a_c}\right)^{\frac{(1-\gamma)}{\gamma}} \tag{9}$$

where $\gamma$ is the ratio of specific heat and $a_c$ is the isentropic constant. For the liquid phase, the relationship between density and pressure is:

$$\rho - \rho_o = \psi(p - p_o) \text{ , where } \psi = \frac{1}{c^2} = \left(\frac{d\rho}{dp}\right)_s \tag{10}$$

where c is the speed of sound, and $\rho_o, p_o$ are the reference density and pressure, respectively. The species transport equation is solved by assuming the liquid phase as a single component system and the gaseous phase as a multi-component system.

$$\frac{\partial(\alpha_2\rho_2Y_k)}{\partial t} + \nabla.(\alpha_2\rho_2\mathbf{u}Y_k) - \nabla.(\alpha_2D_k\nabla Y_k) = S_{\rho Yk}\big|_L + S_{\rho Yk}\big|_E \tag{11}$$

where $Y_k$ and $D_k$ are the mass fraction and coefficient of diffusion for the $k^{th}$ species in the gaseous phase. The energy equation for high-temperature evaporating conditions, formulated by Bhatia et al. [6] is:

$$\frac{\partial(\rho T)}{\partial t} + \nabla.(\rho\mathbf{u}T) - \nabla.(\alpha_{T,eff}\nabla T) + \left(\frac{\partial(\rho K)}{\partial t} + \nabla.(\rho\mathbf{u}K) + \nabla.(\mathbf{u}p)\right)\left(\frac{\alpha_1}{c_{v,1}} + \frac{\alpha_2}{c_{v,2}}\right) = S_{\rho T}\big|_L + S_{\rho T}\big|_E \tag{12}$$

$$S_{\rho T}|_E = S_\rho|_E * L(T_b)/Cp_{mix} \tag{13}$$

$$S_{\rho T}|_L = S_\rho|_L * L(T_b)Cp_{mix} \tag{14}$$

Here, T is temperature, $\alpha_{T,eff}$ is effective thermal diffusivity, K is the kinetic energy, and $c_{v1}$, $c_{v2}$ are specific heat capacities at constant volumes for phases 1 and 2, respectively. $L(T_b)$ is the latent heat of vaporization at boiling temperature, and the terms $S_{\rho T|L}$ and $S_{\rho T|E}$ are used to enforce the saturation condition, which represents energy contributions due to phase change for Eulerian and Lagrangian droplets, respectively. More details on the energy equation and source terms are thoroughly explained in Johny et al.[36]. Although evaporation is negligible under the current low-temperature conditions, the framework is designed to incorporate its influence within both Eulerian and Lagrangian domains.

Turbulence is modeled using a one-equation eddy viscosity subgrid-scale (SGS) model, specifically the LES dynamic k-equation model. The governing partial differential equations are solved numerically using the finite volume method (FVM) in the OpenFOAM-v1912 platform[38]. A second-order total variation diminishing (TVD) scheme is used for convective fluxes, while a second-order central scheme is applied to viscous fluxes. Time integration is performed using a first-order implicit Euler scheme with sufficiently small time steps to ensure numerical stability and minimize numerical diffusion.

### B. Lagrangian Framework

In the LPT method, liquid droplets are treated as point particles possessing mass and momentum but are devoid of volume. To efficiently represent dispersed droplets, the method employs parcels-groups of droplets with similar properties such as size, velocity, temperature, and thermophysical characteristics. The parcel position and velocity are updated iteratively at each time step using the Basset-Boussinesq-Oseen (BBO) equation[40]:

$$\frac{d\mathbf{x}_p}{dt} = \mathbf{u}_p \tag{15}$$

$$m_p \frac{d\mathbf{u}_p}{dt} = \mathbf{F}_D + \mathbf{F}_G \tag{16}$$

where $\mathbf{x}_p$, $m_p$ and $\mathbf{u}_p$ denote the particle position, mass, and velocity, respectively. The drag force $\mathbf{F}_D$ and gravitational forces $\mathbf{F}_G$ (body forces including buoyancy effects) acting on each particle is given by:

$$\mathbf{F}_D = C_D \frac{\pi D_p^2}{8} \rho g(\mathbf{u}_g - \mathbf{u}_p)|\mathbf{u}_g - \mathbf{u}_p| \tag{17}$$

$$\mathbf{F}_G = m_p g \left(1 - \frac{\rho_g}{\rho_p}\right) \tag{18}$$

where $D_p$ is the parcel diameter, and $\mathbf{u}_g$ and $\mathbf{u}_p$ are the velocities of the gas phase and Lagrangian parcels, respectively. The drag coefficient $C_D$ is calculated using the Schiller-Naumann correlation[51]:

$$C_D = \begin{cases} 24(1 + 0.15(Re_p)^{0.687})/Re_p & Re_p \leq 1000 \\ 0.44 & Re_p > 1000 \end{cases} \quad (19)$$

$$\mathrm{Re}_p = \frac{\rho_2|\mathbf{u}_g - \mathbf{u}_p|D_p}{\mu_2} \quad (20)$$

where $Re_p$ is the droplet Reynolds number, and $\mu_2$ is the gas phase dynamic viscosity. The LPT framework models various droplet phenomena, including breakup, heat transfer, and evaporation. Heat transfer calculations utilize the Ranz-Marshall model[44] , while evaporation, assuming spherical Lagrangian droplets, is determined using the Frössling correlation[18]:

$$Sh = 2 + 0.552{Re_p}^{1/2}Sc^{1/3} \quad (21)$$

where the Sherwood number ($Sh$) represents the ratio of the convective mass transfer coefficient ($h_c$) times the droplet diameter ($d_p$) to the characteristic length scale ($D$). The droplet Reynolds number ($Re_p$) and the Schmidt number ($Sc$) significantly influence the evaporation rate. The turbulent Schmidt number (or Prandtl number), which typically ranges from 0.7 to 1, is set to 0.85 in this study.

The secondary breakup of Lagrangian parcels induced by aerodynamic shear forces is modeled using the Reitz-Diwakar breakup model[45,46], wherein droplets exceeding a critical Weber number undergo either bag or stripping breakup, producing smaller droplets from the parent parcels. The breakup frequency is calculated based on droplet diameter, the relative velocity between droplets and gas phase, and specific model constants: $C_{bag}$=6.0, $C_b$=0.78, $C_{strip}$=0.5, and $C_s$=10.0. Additionally, droplet collision and coalescence phenomena are accounted for through the Schmidt and Rutland[52] trajectory collision model.

### C. Coupling Algorithm

The coupling between the compressible VOF and LPT frameworks, facilitating the transition of droplets from Eulerian to Lagrangian, is based on the methodology formulated by Heinrich & Schwarze[21]. The individual droplets in the 3D Eulerian framework are identified using the CCL algorithm. Droplets satisfying predefined size and shape criteria, specifically with a maximum droplet diameter of 280 μm and a maximum sphericity of 1.47, are converted into Lagrangian parcels.[1,2] Droplets exceeding these thresholds are retained in the Eulerian framework, as larger droplets typically exhibit continued deformation and may undergo subsequent breakup downstream. The maximum sphericity cutoff applied here is experimentally determined, serving to effectively exclude ligaments and irregularly shaped droplets from the analysis. The present study adopted a one-way coupling approach, allowing the conversion from Eulerian droplets to Lagrangian parcels but not vice versa.

### D. Domain Specification, Boundary Conditions, and Solution Parameters

The computational model has been previously validated in Bhatia et al.[6]. A 3D computational domain replicating the experimental channel, as depicted in Fig. 1 (a), is employed, with calculations for $D_{32}$ and STD performed in the downstream sub-domain. The liquid jet ($D_n$ = 572 μm, water at 25 °C) is injected into an air cross-flow (25 °C) under a fully developed turbulent inlet profile, while the jet nozzle imposes a uniform velocity profile. Detailed fluid properties and simulation parameters are presented in Tables I and II. Wave-transmissive conditions are applied at the outlet, and all solid walls are treated as no-slip, adiabatic with *k-q-R* wall functions.

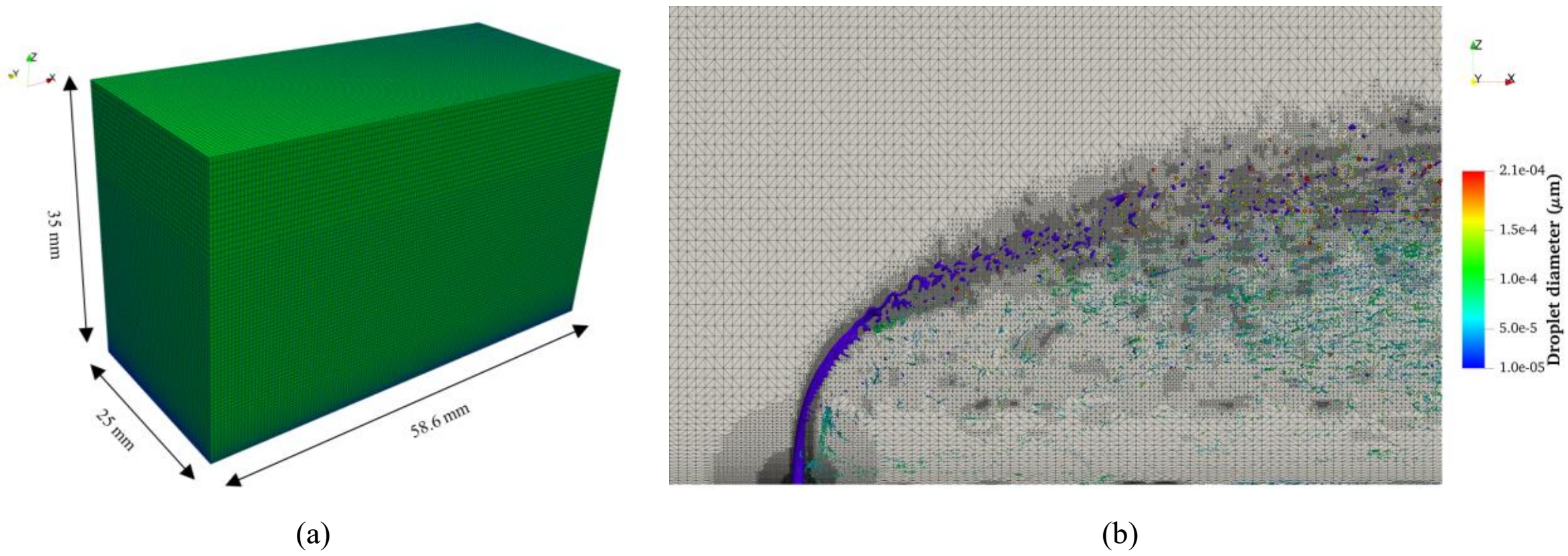

FIG 1. (a) Computational domain meshed with 0.75 million cells (b) Iso-contour value $\alpha$ =0.5 representing Eulerian part (violet), and discrete droplets indicate lagrangian droplets

The base mesh consists of 140 × 60 × 80 cells, refined near the injector using a multi-level refinement approach via *snappyHexMesh*, achieving a resolution ranging from 26-418 μm and yielding a 0.75-million-cell parent mesh. Adaptive mesh refinement enhances interface resolution, while dynamic grid adjustments, including buffer layers, accommodate droplet formation and movement, as shown in Fig. 1 (b). The mesh dynamically adapts to interface evolution, reverting to the base mesh cell size upon Eulerian → Lagrangian conversion to maintain consistency with Lagrangian assumptions. To ensure numerical stability, the simulation operates under a strict Courant–Friedrichs–Lewy (CFL) constraint (CFL < 0.12). Buffer layers near Eulerian droplets facilitate a gradual transition in cell size, preventing numerical instabilities. The total cell count varies dynamically between 0.75 and 3–5 million.

TABLE I. Summary of fluid properties

| Fluid Property | P=2.1 bar, T=25°C | P=3.8 bar, T=25°C |
|---|---|---|
| $\rho_{jet}$ (kg/m$^3$) | 997.10 | 997.17 |
| $\rho_{air}$(kg/m$^3$) | 2.42 | 4.44 |
| $\sigma$ (N/m) | 0.072 | 0.072 |
| $\nu_{air}$(m$^2$/s) | $7.638 \times 10^{-6}$ | $4.167 \times 10^{-6}$ |
| $\nu_{jet}$( m$^2$/s) | $8.927 \times 10^{-7}$ | $8.925 \times 10^{-7}$ |
| $\mu_{air}$(N.s/m$^2$) | $1.849 \times 10^{-5}$ | $1.851 \times 10^{-5}$ |
| $\mu_{jet}$(N.s/m$^2$) | $8.901 \times 10^{-4}$ | $8.900 \times 10^{-4}$ |

Temporal integration employs an implicit first-order Euler scheme with adaptive time-stepping, constrained by a maximum Courant number (CFL < 0.12). Spatial discretisation follows standard OpenFOAM practices: Gauss-linear schemes have been adopted for gradient and Laplacian terms, while bounded, second-order upwind-biased schemes are employed for convection terms. The compressible flow solver uses a PIMPLE algorithm comprising one outer SIMPLE iteration, three inner PISO correctors, and one non-orthogonal correction step. Pressure (p_rgh) is solved with a GAMG solver (tolerance: $10^{-6}$, tightened to $10^{-8}$ for final iteration), velocity with a PBiCGStab solver (tolerance: $10^{-6}$, tightened to $10^{-8}$ for final iteration), and thermodynamic/turbulence scalars with a smoothSolver (tolerance: $10^{-8}$). The volume fraction equation is solved using the isoAdvector geometric scheme with two interface corrections and three sub-cycles per time step. Each simulation runs for 10 ms of physical time, and statistical averaging begins once steady values for trajectory and droplet-size metrics (SMD, $D_{32}$) are observed. Convergence is achieved when all residuals drop below specified tolerances and instantaneous mass imbalance remains below $10^{-5}$.

## III. DROPLET SIZE DISTRIBUTION (DSD) AND KEY PARAMETERS

The following section outlines these key parameters, relevant terminologies, and distribution functions used for evaluating DSD in spray applications. DSD characterizes the range of droplet sizes within a spray and is typically represented by histograms or frequency distribution curves detailing droplet count, surface area, or volume across different size ranges[28]. Volume-based distributions tend to skew towards larger droplets,[36] as they contribute more significantly to overall volume, as depicted in Fig. 2. Analyzing DSDs helps understand the effects of operating conditions, such as air velocity in atomizers, and aids in optimizing specific applications, from combustion to agricultural spraying. In spray analysis, mean diameters are essential metrics for characterizing droplet size distributions and influencing calculations in mass transfer, evaporation, and combustion.[12]

Various mean diameters are used to characterize droplet size distributions, and each provides unique insights into spray behavior suited for different applications. The *Arithmetic Mean Diameter ($D_{10}$)* provides the linear average of all droplet diameters, making it useful for basic size comparisons when droplet count is prioritized over mass or volume. The *Sauter Mean Diameter ($D_{32}$ or SMD)*, defined as the ratio of particle volume to surface area, provides insights into atomization efficiency. Smaller $D_{32}$ values indicate finer droplets, facilitating faster evaporation and improved mixing, making it particularly valuable in evaporation and combustion applications. As noted by Chin and Lefebvre,[12] $D_{32}$ is often considered the best indicator of spray fineness, especially in assessing spray performance. The expression for mean diameters can be expressed in general as:

$$D_{mn} = \left[\frac{\sum(N_i D_i^{\,m})}{\sum(N_i D_i^{\,n})}\right]^{\frac{1}{m-n}} \tag{22}$$

where $D_i$ is the diameter of each droplet in the distribution, $N_i$ is the number of droplets of diameter $D_i$, $m$ and $n$ are the exponents that vary with respect to the mean diameter taken into consideration.

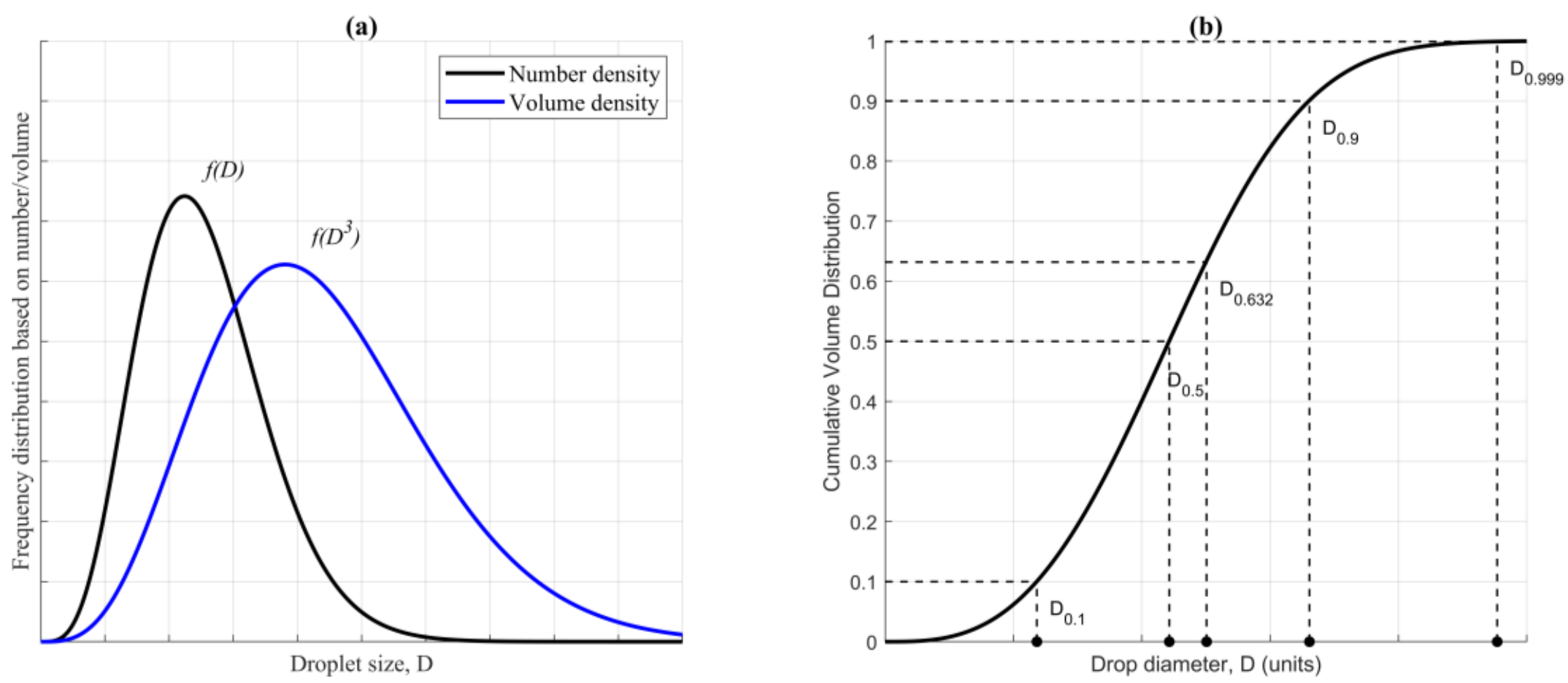

FIG 2. (a) Frequency distribution curve based on number and volume (b) Cumulative volume distribution depicting representative diameters ($D_{0.1}$, $D_{0.5}$, $D_{0.632}$, $D_{0.9}$, $D_{0.999}$)

Different modeling approaches are employed to fit droplet size distributions in spray atomization. *Droplet size distribution (DSD)* functions serve as an essential statistical tool for characterizing and predicting droplet behavior. The probability density function (PDF), often represented as a number distribution *f(D)*, indicates the likelihood of finding droplets within specific diameter ranges, providing a continuous representation of the droplet size distribution.[28] In addition, volume distributions *$f(D^3)$* illustrate the fraction of spray volume contributed by droplets of various sizes, which is especially relevant for applications requiring mass and volume control, such as fuel injection. To simplify the complexity of polydisperse sprays, representative diameters such as $D_{10}$ (arithmetic mean), $D_{30}$ (volume mean), and $D_{32}$ (Sauter mean diameter) are calculated from PDF moments, offering insights into surface area or volume-to-surface ratios that characterize spray behavior.[36] For JICF sprays, probability density plots (PDP) can reveal bimodal distributions, often resulting from secondary breakup processes. The cumulative distribution plot (CDP) complements the PDP by showing the cumulative percentage of droplets by number/volume below a certain diameter, aiding in the analysis of droplets within target size ranges for specific applications. Discrete size distributions (histograms) are plotted here using round symbols at the center of each bin width, which are used to obtain continuous size distributions. All the additional parameters, including statistical measures such as the mean diameters, dispersion parameters, and distribution functions, have been included in the appendix for reference.

## IV. RESULTS AND DISCUSSION

In this study, the droplet distribution obtained for a range of test cases is analyzed for a LJICF modeled using the VOF-LPT coupled framework. Tables I and II show the fluid properties and test conditions employed in the simulations, respectively. Further sections deal with how the crossflow and liquid jet parameters affect the droplet size distribution and transport downstream of jet injection. Since a coupled framework is used for the spray atomization simulations, which seamlessly transforms a range of Eulerian droplets into lagrangian, it is imperative to analyze the resulting droplet size distributions obtained. The calculation of droplet size characteristics involves identifying the droplet size and position for each Lagrangian droplet, categorizing them into respective bins, and then averaging over time to generate the mean droplet distribution plots. The calculations for mean droplet sizes ($D_{32}$, $D_{10}$, STD) and droplet size distribution have been employed by taking an average over time within the specific regions of the domain.

### A. DROPLET SIZE CHARACTERISTICS AND VALIDATION

The Eulerian-Lagrangian coupled framework employed in this study was previously formulated and validated by Bhatia et al. [6] in terms of liquid jet penetration and mean droplet sizes. In this work, an extended analysis is conducted, exploring a wider range of parameters to gain deeper insights, specifically into DSD and transport characteristics. Figure 3 presents scatter plots of $D_{32}$ and STD versus momentum flux ratio, comparing simulation results with experimental data for droplet characteristics from the breakup of a liquid jet in crossflow. A ±10% error band is shown on the CFD results due to the absence of statistical uncertainty data from the experiments. While the simulations generally align well with experimental data, variations in the Weber number across the scatter points make it challenging to establish a definitive trend. The isolated effects of the momentum flux ratio and Weber number are addressed in later sections.

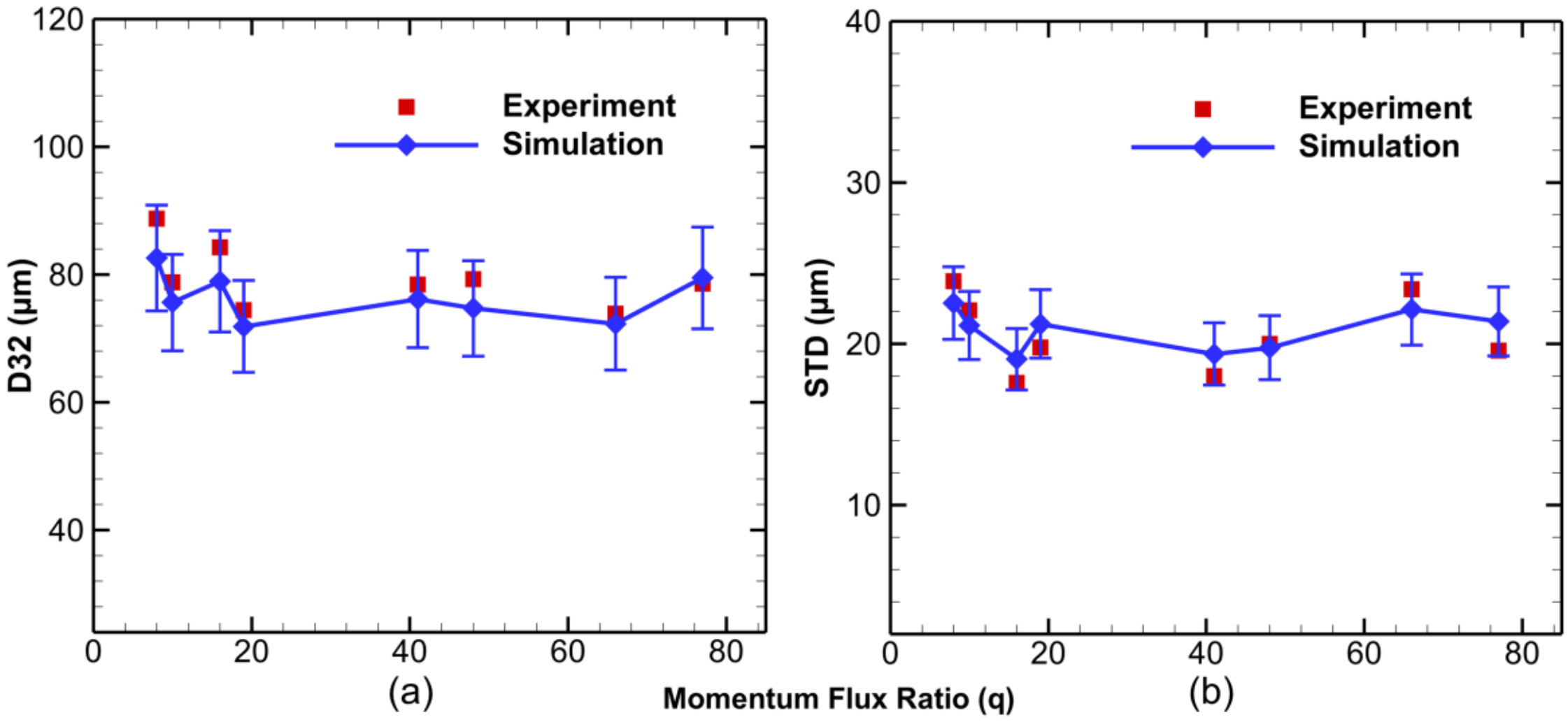

FIG 3. (a) Scatter plots of Sauter mean diameter ($D_{32}$) and (b) standard deviation (STD) against experimental data for different momentum flux ratios (q). Each data point here represents the mean droplet size obtained for individual cases within the 52.45D downstream of injection, which is taken as per the experiments.

TABLE II. Comparison of droplet size characteristics for computational and experimental cases.

| *Case No.* | **Pressure (bar)** | **Crossflow velocity (m/s)** | *Jet velocity (m/s)* | *Momentum flux ratio (q)* | *Weber number (We)* | *Experimental* | | *Computational* | | |
|---|---|---|---|---|---|---|---|---|---|---|
| | | | | | | *$D_{32}$ (μm)* | *STD (μm)* | *$D_{32}$ (μm)* | *STD (μm)* | *Error in $D_{32}$ (%)* |
| 1 | 2.1 | 61 | 9 | 8 | 71 | 88.8 | 23.9 | 82.64 | 22.54 | 6.94 |
| 2 | 2.1 | 61 | 12 | 16 | 71 | 84.3 | 22.1 | 78.97 | 21.14 | 6.32 |
| 3 | 2.1 | 61 | 19 | 41 | 71 | 78.5 | 19.6 | 76.18 | 20.61 | 2.96 |
| 4 | 2.1 | 61 | 24 | 66 | 71 | 73.9 | 17.6 | 72.34 | 19.06 | 2.11 |
| 5 | 3.8 | 65 | 14 | 10 | 150 | 78.8 | 19.8 | 75.65 | 21.24 | 4.00 |
| 6 | 3.8 | 65 | 19 | 19 | 150 | 74.5 | 18.0 | 71.90 | 19.38 | 3.49 |
| 7 | 3.8 | 41 | 19 | 48 | 60 | 79.3 | 20.0 | 74.75 | 19.77 | 5.74 |
| 8 | 3.8 | 41 | 9 | 10 | 60 | 87.6 | 23.4 | 82.84 | 22.14 | 5.43 |
| 9 | 3.8 | 41 | 24 | 77 | 60 | 75.2 | 18.1 | 68.96 | 17.42 | 8.30 |
| 10 | 3.8 | 33 | 19 | 77 | 38 | 78.6 | 19.6 | 79.52 | 21.39 | -1.17 |
| 11 | 3.8 | 65 | 30 | 48 | 150 | 70.7 | 16.2 | 65.18 | 17.17 | 7.81 |
| 12 | 3.8 | 65 | 38 | 77 | 150 | 66.7 | 14.2 | 65.62 | 18.11 | 1.62 |
| 13 | 3.8 | 41 | 12 | 19 | 60 | 84.1 | 22.0 | 78.52 | 20.73 | 6.63 |
| 14 | 3.8 | 92 | 19 | 10 | 298 | 69.5 | 15.6 | 61.39 | 16.67 | 11.67 |
| 15 | 3.8 | 92 | 27 | 20 | 298 | 66.9 | 14.3 | 61.85 | 15.77 | 7.55 |
| 16 | 3.8 | 92 | 43 | 49 | 298 | 65.2 | 13.5 | 60.48 | 18.70 | 7.24 |

Table II shows the set of cases analyzed in this particular study, which is based on the experimental results of Amighi and Ashgriz.[2,3] All the simulations were performed at a constant temperature of 25ºC and under high-pressure conditions, specifically at 2.1 and 3.8 bar, to mimic the density ratios as in actual gas turbines. The tests covered a range of momentum flux ratios (from 8 to 77) and crossflow Weber numbers (from 38 to 298). The number of generated droplets also varied depending on the test conditions, ranging from 40,000 to 8,00,000, reflecting the extent to which the liquid jet core penetrated and broke up in each scenario. The comparison of droplet sizes against the experiments showed an average error of 5.6 % and 9.6 % for $D_{32}$ and STD, respectively. Hence, the obtained data in simulations for droplet characteristics are found to be in good agreement with the experimental data. The $D_{10}$ obtained in the simulations is found to be lower than that in the experiments. An average error of 12.35% is obtained for $D_{10}$. The observed lower $D_{10}$ in the simulation is attributed to methodological differences between experimental and numerical droplet size calculations. Specifically, the simulations account for droplets undergoing breakup into substantially smaller sizes, whereas in the referenced experiments, very small droplets are not captured and are therefore excluded from size calculations. Conversely, the predicted $D_{32}$ aligns closely with experimental results, as larger droplets predominantly influence $D_{32}$ and are thus less sensitive to the presence of very small droplets. To closely replicate the experimental measurement approach and accuracy, our numerical analysis considered only those Lagrangian droplets converted from the Eulerian framework, ensuring the inclusion of droplets exhibiting relevant size characteristics as mentioned in section II.B, conforming to the experiments.

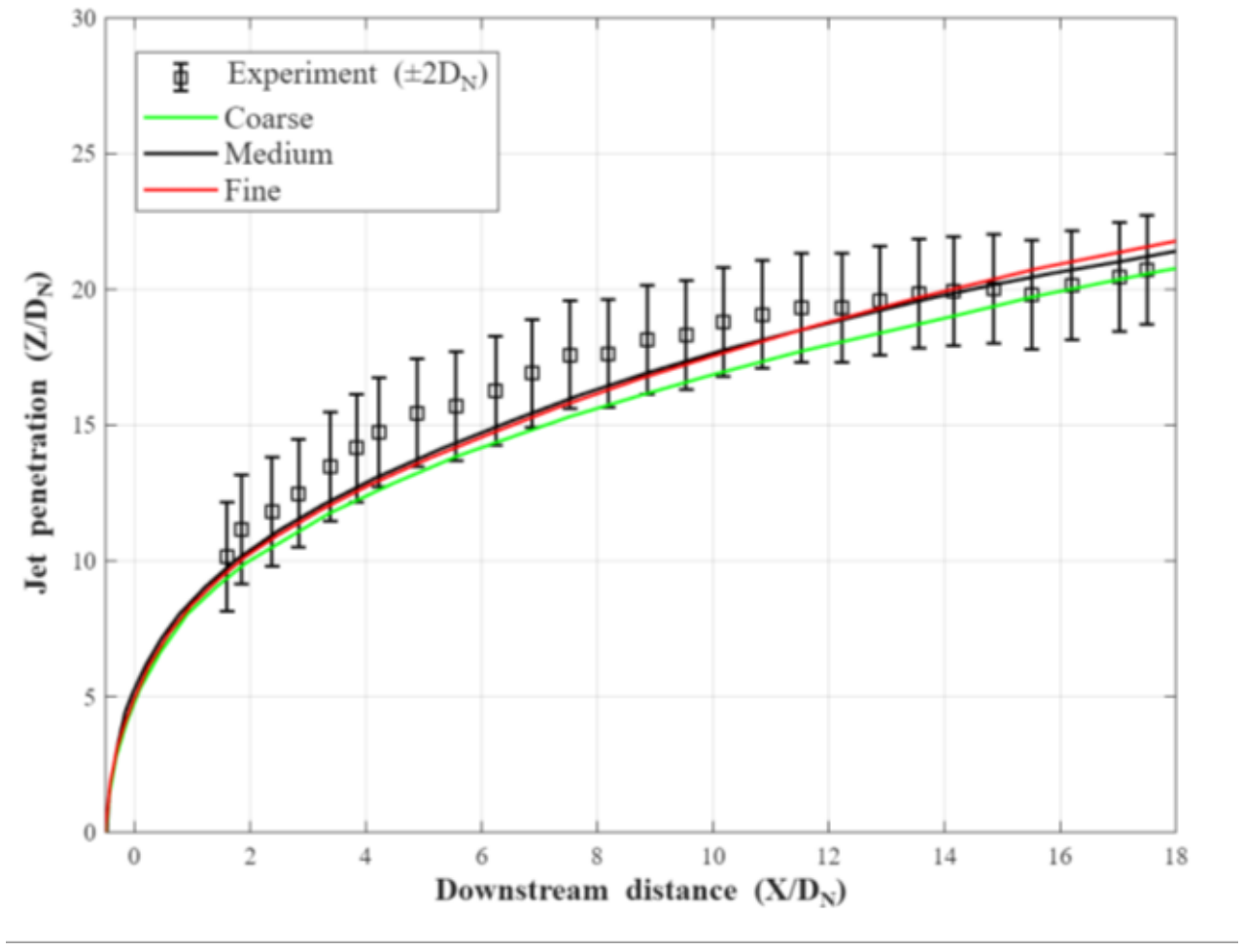

FIG 4. Windward jet trajectories for Case 6 on coarse (~1.3 M), medium (~4 M) and fine (~7 M) meshes versus experiment, demonstrating mesh independence.

A grid-sensitivity test is also performed for Case 6 using three progressively refined meshes, coarse (≈1.3 M cells: $D_{32}$ = 78.75 µm, STD = 21.62 µm), medium (≈4 M cells: $D_{32}$ = 71.90 µm, STD = 19.38 µm) and fine (≈7 M cells: $D_{32}$ = 71.07 µm, STD = 20.08 µm) to assess mesh independence in both windward trajectories and droplet size metrics. As illustrated in Fig. 4, all three grids capture the jet trajectory within experimental uncertainty, with only a slight underprediction in the coarser mesh. For the medium and fine grids, the trajectories coincide well until 15D, and later on, only a slight deviation is observed. The minimal variation in near-nozzle trajectories is attributed to the efficacy of the adaptive mesh refinement (AMR) technique, which ensures adequate resolution of interface dynamics even on coarser base grids. Moreover, the $D_{32}$ values obtained on the medium mesh deviate by less than 1 μm (<2 %) from those of the fine mesh, confirming convergence of the droplet size metrics.

## B. EFFECT OF CROSSFLOW AND JET PARAMETERS ON OVERALL DROPLET SIZE DISTRIBUTION

The overall mean droplet size characteristics in Table II were calculated in the 52.45D downstream domain from the liquid jet injection point as per the experimental calculations. The number-based probability density plot (PDP) for a jet-in-crossflow (LJICF) categorizes the droplet sizes produced during atomization in different size ranges as a fraction of the total number of droplets in the domain. The cumulative distribution plot (CDP) based on the number or volume of droplets represents the proportion of total droplets (by count or volume) that are smaller than or equal to a given size. This helps to visualize the spread of droplet sizes within a spray and assess how they contribute cumulatively to the distribution.

### B.1 EFFECT OF MOMENTUM FLUX RATIO

Figures 5 (a) and (b) show the number based probability density plot (PDP), and Fig. 5 (c) and (d) show the volume based cumulative distribution plot (CDP) of droplet sizes at different liquid jet velocities and two different pressure conditions (2.1 bar and 3.8 bar). Figures 5 (a) and b reveal that droplets above 100 μm are also present, though few in number, and contribute significantly to $D_{32}$, as the CDP shows that 20-40% of the contribution of volume is from droplets above 100 μm. This highlights that the number based PDP alone doesn't capture the influence of larger droplets on $D_{32}$ and STD. Therefore, a volume-based PDP or CDP is more suitable for the qualitative analysis of these distributions.

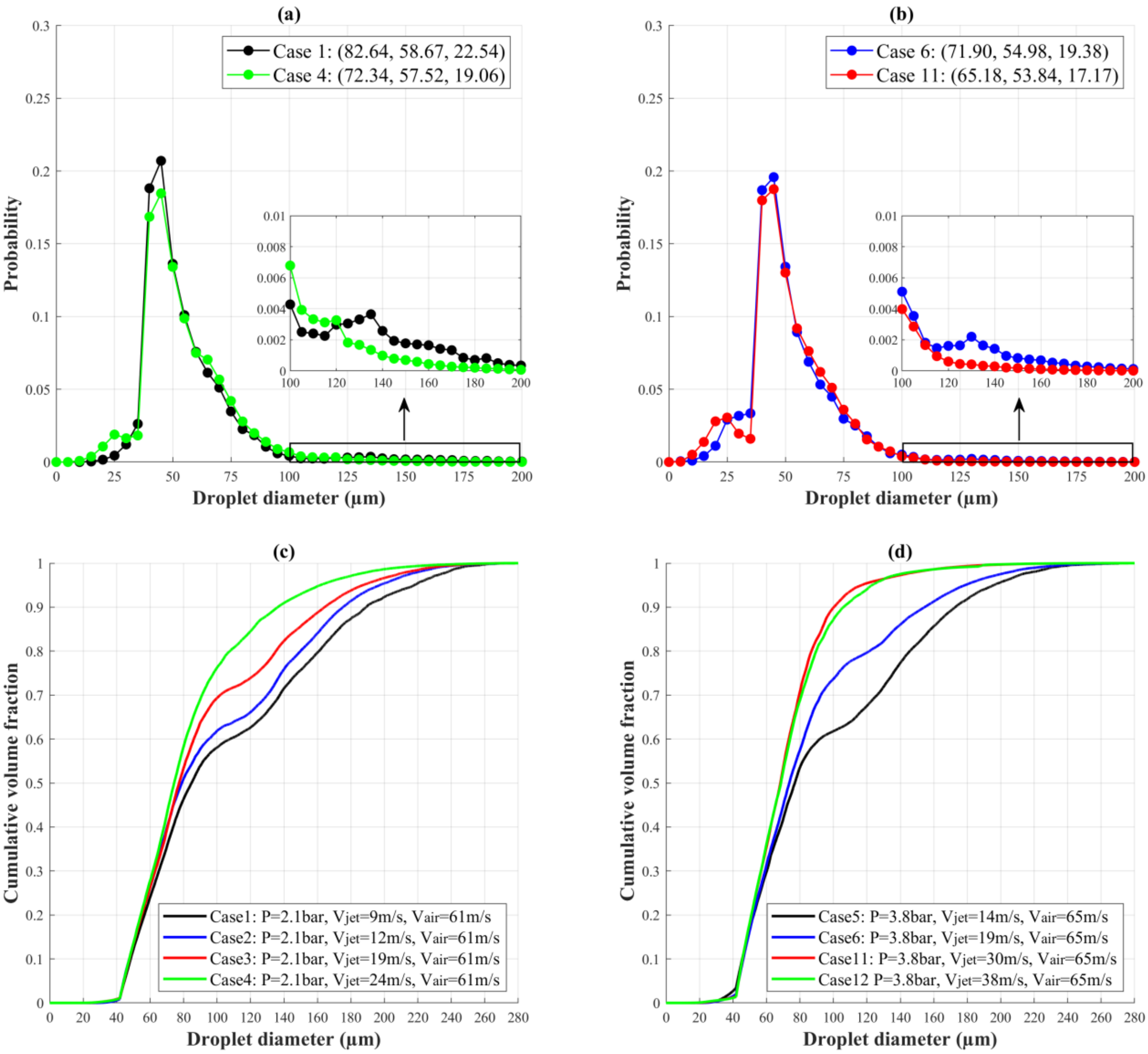

FIG 5. Effect of liquid jet velocity on (a), (b) overall droplet size distribution (PDP) and (c), (d) cumulative droplet size distribution (CDP). Legend values in the brackets represent ($D_{32}$, $D_{10}$, STD)

As the liquid jet velocity increases, the cumulative distribution curves in Fig. 5 (c) and (d) shift to the left, indicating smaller average droplet diameters. At 2.1 bar, increasing the jet velocity from 9 m/s to 24 m/s results in a steeper CDP curve that peaks at a smaller diameter. The higher liquid jet velocities result in an increased momentum flux ratio and consequently induce the Rayleigh-Taylor instability in the jet core to produce smaller droplets.[7] A similar trend appears at 3.8 bar, though with a shift toward smaller droplet sizes than at 2.1 bar, highlighting the role of both pressure and velocity in droplet breakup. Higher pressures (3.8 bar) with large jet velocities (30 and 38 m/s) lead to steeper slopes at smaller diameters, indicating efficient atomization with a more uniform DSD, which could be advantageous for applications requiring fine sprays, such as fuel-air mixing. Thus, an increased momentum flux ratio enables the cumulative volume to be reached with smaller diameter droplets, implying a more efficient atomization process that could enhance fuel-air mixing or similar applications in spray technology. For cases 11 and 12, the CDP exhibits similar behavior due to the high q, which causes significant jet penetration and top-wall impingement for larger droplets in case 12. Additionally, the R-T wavelength is closer, depicting similar atomization levels. While q increases from 48 (Case 11) to 77 (Case 12), the Weber number remains unchanged. This implies that increasing $V_{jet}$ in this case does not significantly enhance the breakup process.

The experimental $D_{32}$ decreases slightly from 70.7 µm (Case 11) to 66.7 µm (Case 12). However, the computational $D_{32}$ values for both cases are almost identical (65.18 µm for Case 11 and 65.62 µm for Case 12), indicating that computational predictions are less sensitive to the increase in jet velocity from this point. This is attributed to the lower cutoff (42 µm) conversion criterion used in the simulations for Eulerian-to-LPT transformation, which restricts the generation of very small droplets. This cutoff is defined based on the range of droplet sizes analyzed in the experiments.[2,3] Additionally, within the LPT framework, the droplets can further break up into even smaller sizes. The STD values (indicative of the spread in droplet size distribution) show minimal change between the two cases, further supporting the similarity in atomization behavior.

The lower standard deviation at higher velocities also suggests that higher jet velocities contribute to a more uniform DSD. There is also a significant increase in the production of smaller droplets at these higher velocities, with the peak of the PDP reflecting the most common droplet size being noticeably less at lower velocities. This shift occurs because the PDP shows the fraction of total droplets rather than their number density, as illustrated in the subsequent number density plots. Overall, this confirms a trend toward finer and more uniform droplet sizes with increased momentum flux.

### B.2 EFFECT OF CROSSFLOW PRESSURE AND CROSSFLOW VELOCITY

The PDP plots (Fig. 6 (a) and (b)) show that as crossflow pressure or velocity increases, the DSD shifts towards smaller diameters. This indicates that both higher pressures and velocities promote droplet breakup, leading to finer droplets. Cases with higher crossflow pressure or velocity exhibit peaks at smaller droplet sizes, demonstrating more intense atomization effects. This DSD shift reflects the resulting variations in $D_{32}$, $D_{10}$, and STD values, with smaller droplet diameters and reduced spread in droplet sizes at higher crossflow conditions.

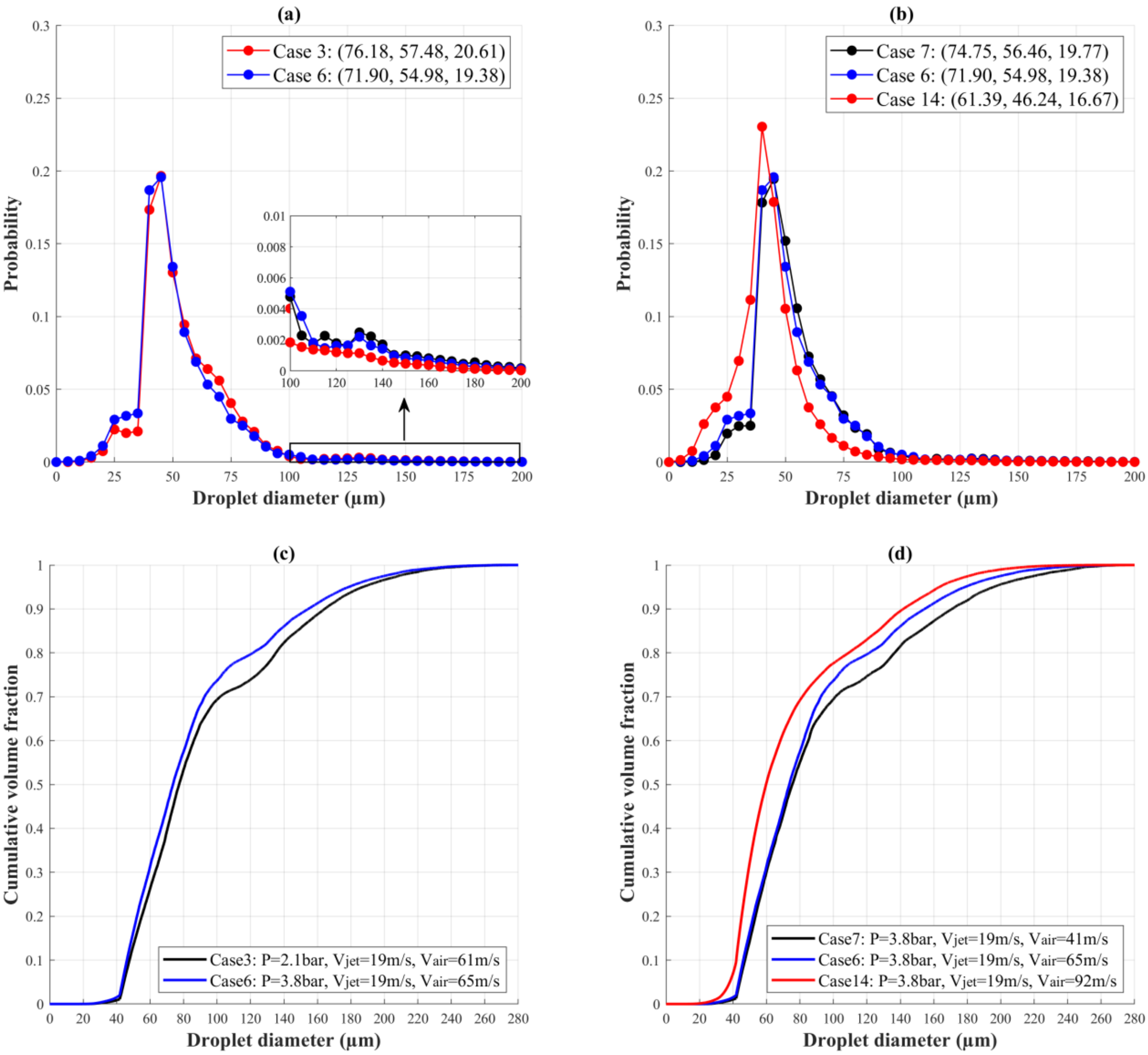

FIG. 6. Effect of crossflow pressure (a), (c) and crossflow velocity (b), (d) on overall droplet size distribution (PDP) and cumulative droplet size distribution (CDP)

The CDP plots (Fig. 6 (c) and (d)) further highlight that increased crossflow pressure and velocity result in a steeper rise in cumulative volume at smaller droplet sizes, indicating that a significant portion of the volume is contained in smaller droplets under these conditions. Higher pressures and velocities cause more efficient atomization, achieving the cumulative volume with finer droplets. The observed trend suggests that controlling crossflow parameters can effectively tailor the droplet size distribution for applications requiring specific droplet characteristics, such as enhanced mixing or controlled spray patterns.

## B.3 EFFECT OF CROSSFLOW WEBER NUMBER

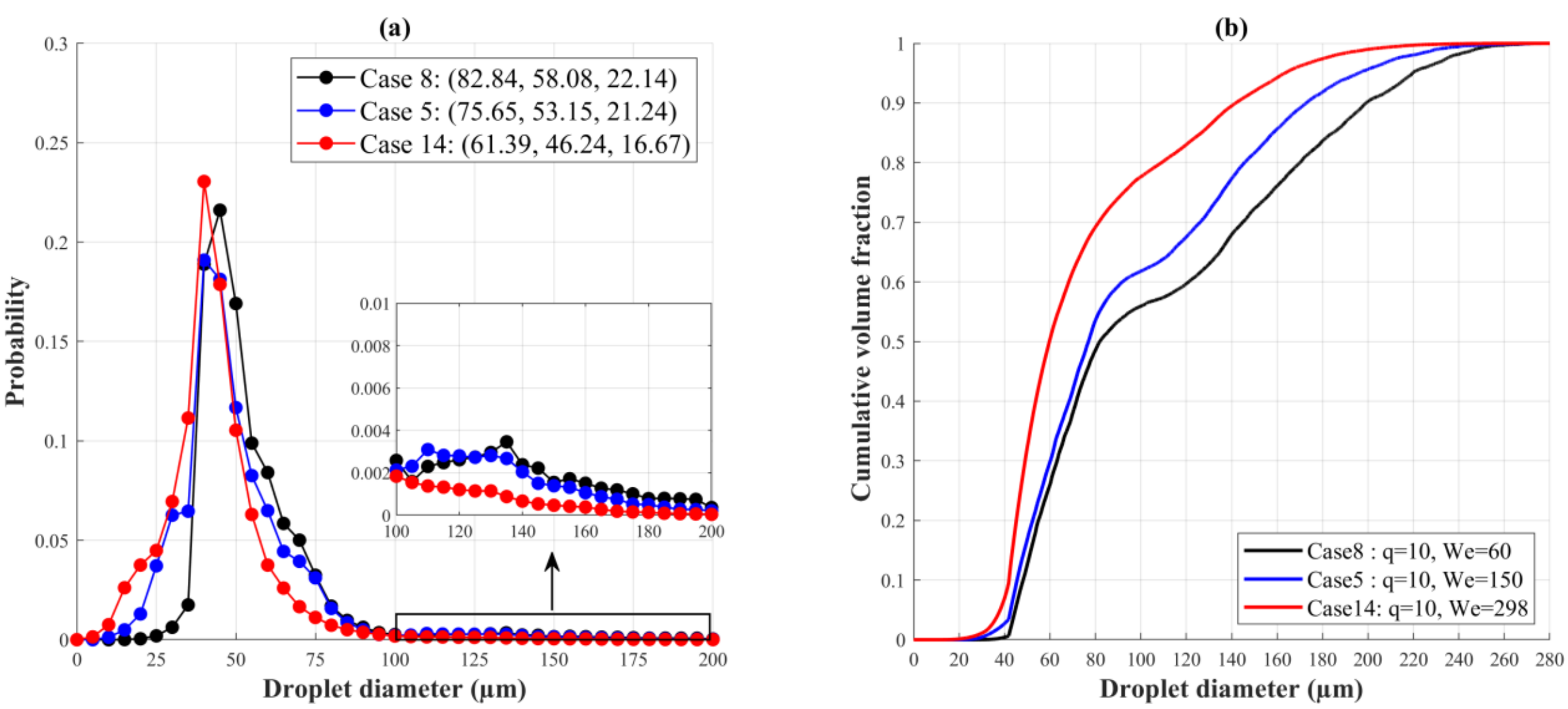

FIG. 7. Effect of crossflow Weber number on (a) overall droplet size distribution (PDP) and (b) cumulative droplet size distribution (CDP)

Figure 7 shows the probability plots for droplet size distribution and cumulative volume fraction for cases 8, 5, and 14, illustrating the effect of Weber number on droplet formation. In the PDP plot, Case 8 (We=60) shows a broader distribution with a higher peak near larger diameters (~45 μm), suggesting a less effective breakup compared to Cases 5 and 14, which have higher Weber numbers (We=150 and We=298, respectively). The higher Weber numbers in Cases 5 and 14 promote more extensive atomization, as reflected by the tighter and higher peaks near smaller diameters (~35-40 μm), indicating a greater prevalence of smaller droplets. The shift observed in the distributions in Case 14 is more pronounced than in Cases 8 and 5, as clearly seen in the PDP. This is because of the increased Weber number, which results in enhanced breakup due to shear. The transition in breakup regimes from column breakup to bag, multimode, and shear with increasing Weber number leads to progressively finer atomization, favoring smaller droplets at higher Weber numbers.

The cumulative volume fraction graph further elucidates these findings. Case 8, with the lowest Weber number, shows the slowest rise in cumulative volume fraction, indicating a distribution skewed towards larger droplets. In contrast, Cases 5 and 14, particularly Case 14 with the highest Weber number, demonstrate steeper curves, reaching 50% of the volume at smaller diameters, which signifies a shift towards finer droplets. This showcases the inverse relationship between the Weber number and droplet size, where higher Weber numbers facilitate the creation of smaller droplets through more efficient atomization and breakup processes.

## C. DROPLET SIZE CHARACTERISTICS VARIATION IN THE STREAMWISE DIRECTION

This section investigates the influence of liquid jet velocity ($V_{jet}$), crossflow pressure (P), and crossflow velocity ($V_{air}$) on droplet size characteristics and their variation downstream. The effects of these crossflow and liquid jet parameters are analyzed, both individually and also through non-dimensional parameters, namely momentum flux ratio (q) and Weber number (We). Subsequent sections delve into the specific impacts of liquid jet velocity, crossflow velocity, and ambient pressure on droplet size distribution using PDP and CDP distributions.

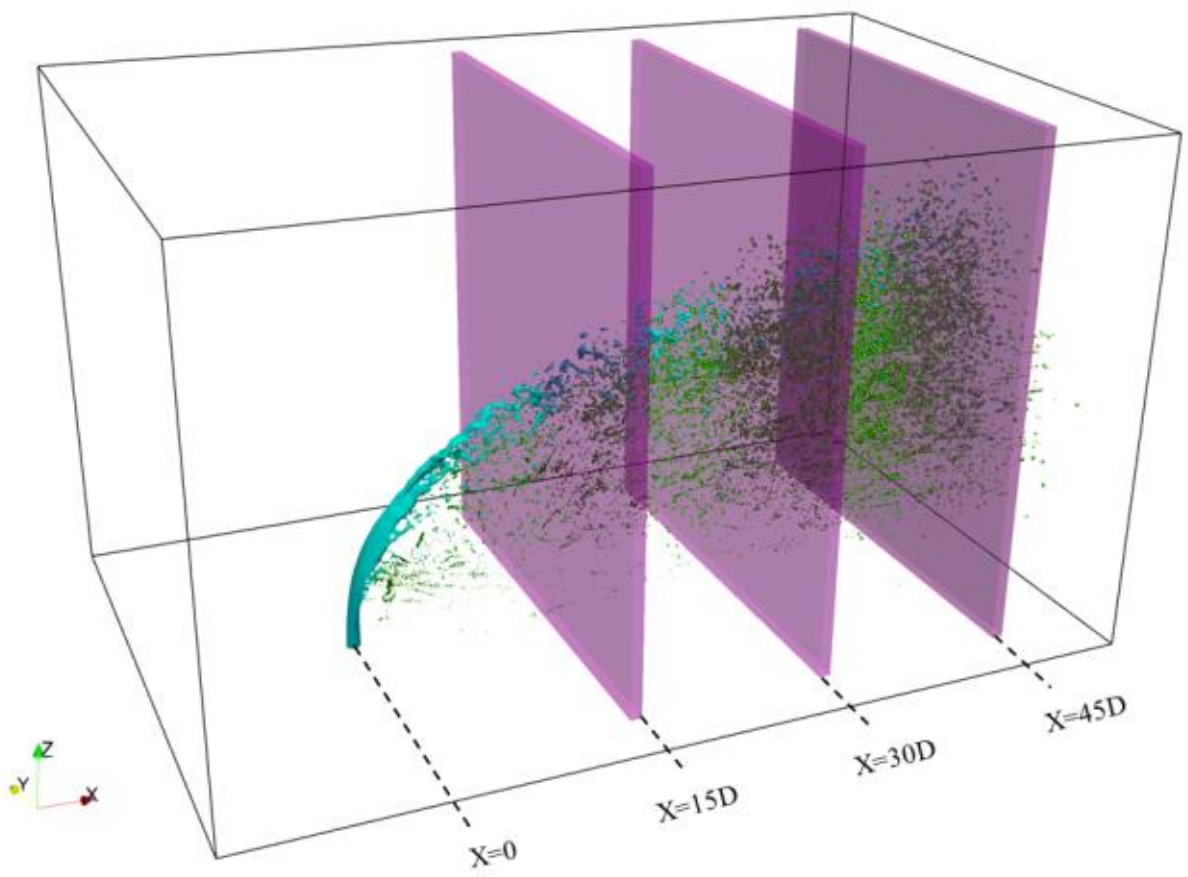

FIG 8. Downstream planes used for the extraction of droplet statistics and size distribution.

Droplet size characteristics were also calculated at downstream locations (X = 5D, 10D, 15D....50D) using 2.5D thick sub-domain planes, as shown in Fig. 8, to understand the variation of droplet characteristics in the downstream. Droplets crossing the downstream planes were recorded and averaged over time to determine mean droplet size characteristics and distribution.

### C.1 EFFECT OF MOMENTUM FLUX RATIO

Figure 9 presents the influence of momentum flux ratio/liquid jet velocity ($V_{jet}$) on droplet size characteristics variation downstream, specifically the $D_{32}$ and the STD of droplets. Across all cases, $D_{32}$ increases up to around 30D downstream and then stabilizes or decreases slightly, reflecting droplet growth and dispersion, as observed from the downstream variation plots. A similar trend has also been observed in Johny et al. [25]. The variation in $D_{32}$ for lower liquid jet velocity cases exhibits a distinct peak and subsequent drop, which can be explained as follows. An initial increase in $D_{32}$ is observed in the near-nozzle region, followed by a slight decrease for lower jet velocity cases, as compared to the higher jet velocity cases that show a steady variation downstream. The initial rise in $D_{32}$ is attributed to the transformation of larger Eulerian droplets into Lagrangian particles, leading to an increase in the volume-to-surface area ratio. In the region close to the nozzle (5–10D), the droplet population predominantly consists of smaller Eulerian droplets sheared from the edges of the liquid jet core in all cases. Moving further downstream, at around 10D, the central liquid jet core begins to disintegrate into larger lumps and ligaments of liquid, depending on the type of instability on the liquid jet surface.[7] These unstable structures break into large droplets between 10D and 20–25D, introducing larger droplets into the Lagrangian system after the conversion[7], as observed from case 1 in Fig. 9 (a) and case 5 in Fig. 9 (c). For very low jet velocity cases, the shearing process at the jet-core edges is far less than that of higher velocity cases. Therefore, more liquid remains as part of the liquid jet core, which undergoes disintegration to produce larger liquid lumps. In other words, a larger amount of liquid in the jet core leads to droplet formation through shearing, in contrast to instability-based central core disintegration. Since these large liquid lumps of low velocity cases are more prone to secondary atomization than the smaller, stable droplets generated from high-velocity cases, the overall droplet size shows a gradual decline in these regions (35D-45D) of low liquid jet velocity cases.

Case 1 (lowest jet velocity) exhibits the highest $D_{32}$, indicating larger droplets, while Case 4 (highest jet velocity) shows the smallest $D_{32}$, reflecting finer droplets due to enhanced atomization at higher jet velocities. A similar trend is seen with STD, where Case 1 exhibits the highest STD, meaning greater size variation in droplets, and Case 4 has the lowest STD, showing more uniform droplet sizes. The decrease in both $D_{32}$ and STD with increasing jet velocity is due to enhanced atomization, driven by a higher momentum flux ratio, as jet velocity increases from 9 m/s to 24 m/s. This results in smaller, more uniform droplets. The observed decline in $D_{32}$ beyond 30 D, even as some droplets > 100 µm persist at 45 D (Fig. 10c), stems from competing droplet dynamics. Initially (10–25 D), core breakup generates large ligaments that convert into bigger droplets, raising $D_{32}$. Beyond 30 D, these large droplets undergo delayed secondary atomization under aerodynamic shear, reducing the mean size, while a small fraction with sufficient inertia survives, accounting for the heavy-end tail at 45 D. Lower jet velocities exacerbate this effect by producing fewer small shear-stripped droplets and more jet core derived lumps that break up downstream.

As the liquid jet velocity increases (Cases 1-4 and Cases 5,6,11), the $D_{32}$ decreases, indicating that higher velocities lead to smaller droplet sizes. This is likely due to stronger atomization forces breaking up the liquid jet more effectively at higher velocities. The larger droplets in Case 1 and Case 5 suggest weaker atomization at lower velocities. Higher liquid jet velocities result in a narrower droplet size distribution (lower STD). This indicates that higher velocities not only produce smaller droplets but also more uniform droplet sizes. The broad distribution of Case 1 (Fig. 9 (a) and (b)) and Case 5 (Fig. 9 (c) and (d)) with lower jet velocities suggests less efficient breakup, leading to a wider range of droplet sizes.

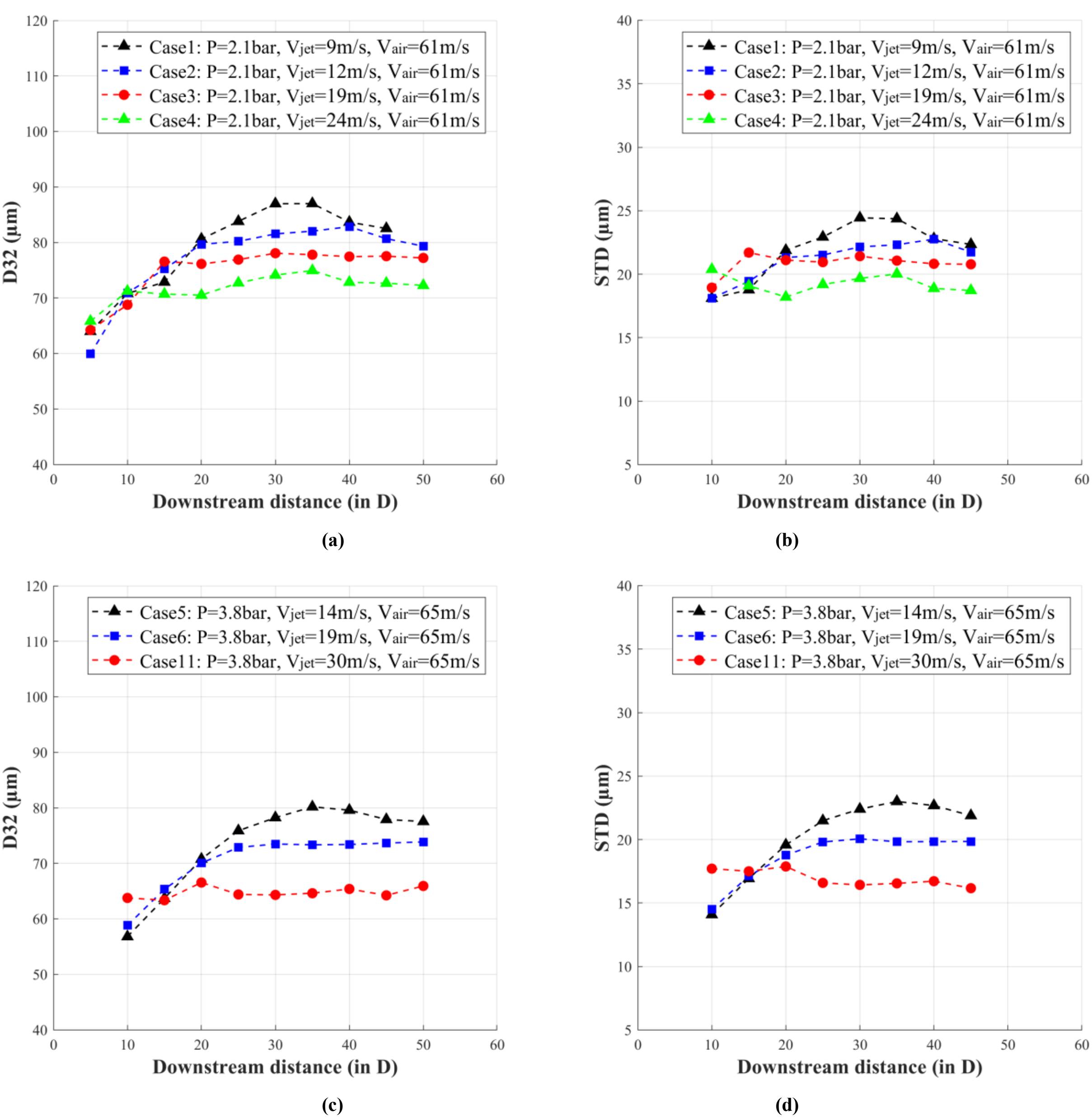

FIG 9. Effect of liquid jet velocity on droplet size characteristics in the downstream at (a), (b) 2.1bar and (c), (d) 3.8bar

At 3.8 bar (Fig. 9 (c) and (d)), the impact of increasing liquid jet velocity on $D_{32}$ and STD remains consistent with the 2.1 bar case (Fig. 9 (a) and (b)): higher velocities result in smaller, more uniform droplets. In Case 11, the absence of a distinct peak and drop trend and the higher $D_{32}$ at 10D compared to Cases 5 and 6 are attributed to the combined effects of high pressure and jet momentum, enabling faster atomization near the nozzle. Minimal CDP variation at 15D, 30D, and 45D (Fig. 11 (d)) confirms that atomization is nearly complete by 15D, with most droplets forming between 10D and 15D. This suggests that the majority of the droplets are formed between 10D and 15D from the injection point, resulting in steady $D_{32}$ values further downstream. The combination of high pressure and jet momentum, as in Case 11, produces the smallest $D_{32}$ and STD, driven by stronger shear forces that break the jet into finer, more consistent droplets.

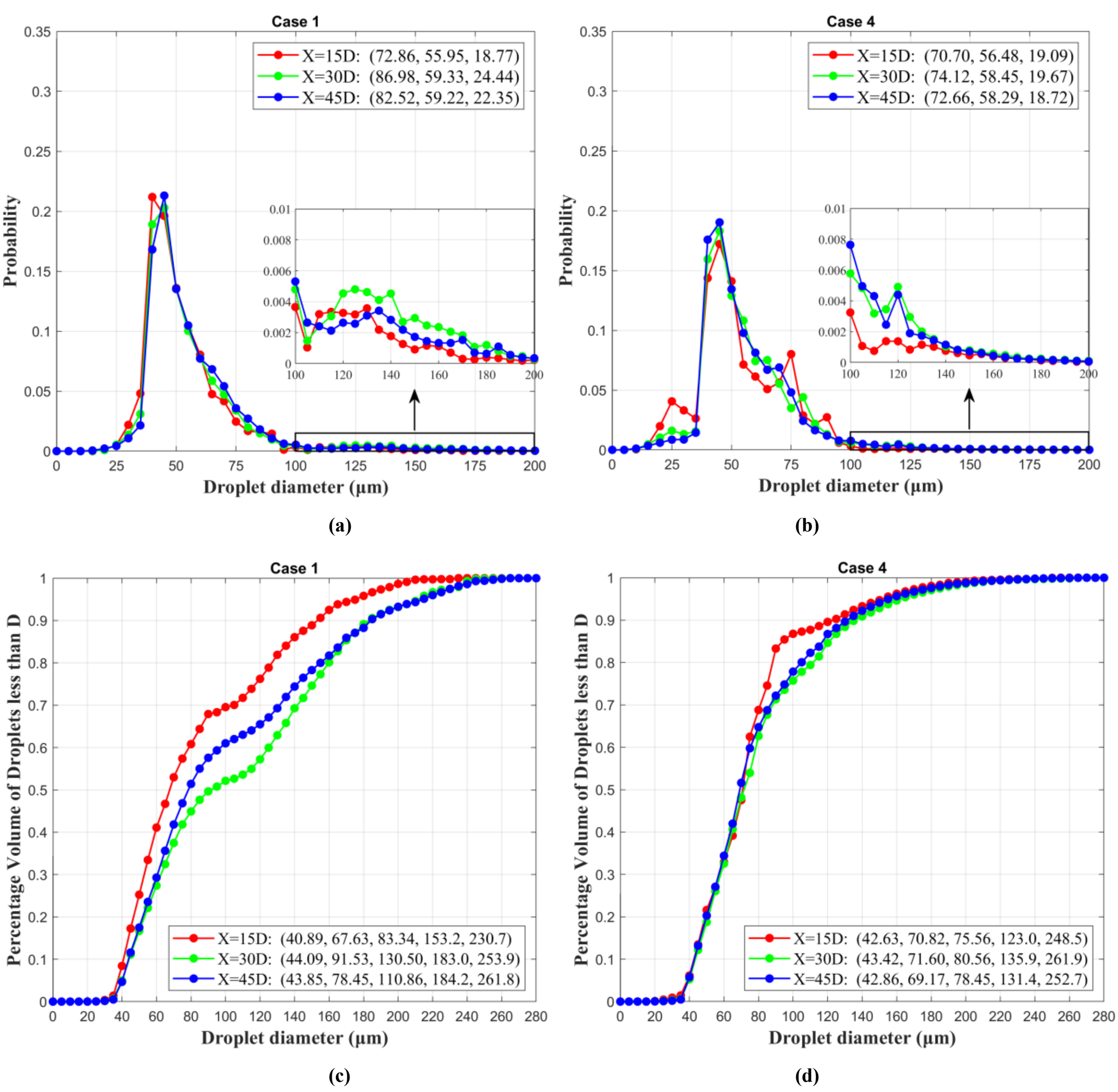

FIG 10. Droplet size distribution in the downstream: (a) Case 1-PDP (b) Case 4-PDP (c) Case 1-CDP (d) Case 4-CDP

Two key factors contribute to the reduction in droplet size. Firstly, a rising jet velocity increases the jet's Reynolds number, leading to higher instability,[30,39] promoting more extensive droplet break-up. Secondly, higher jet velocity enhances penetration depth into the cross-flow, resulting in increased exposure of the liquid jet to the crossflow air. Thus, the increased shear force exposure between the liquid jet and crossflow air leads to finer and more uniform atomization. Thus, a higher atomization efficiency is observed at high liquid jet velocities**.** The combined effect of these factors is a decrease in both $D_{32}$ and STD values. The plot at 3.8 bar reinforces the trend that higher jet velocity and higher pressure both contribute to better atomization, resulting in smaller and uniform droplets. In conclusion, Figs 9 (a), (b), (c), and (d) together demonstrate that increasing both pressure and jet velocity can significantly improve atomization efficiency in spray processes.

Figure 11 presents the PDPs and CDPs at a slightly higher pressure of 3.8bar for cases 6 and 11 in the downstream locations of 15D, 30D, and 45D. Some general observations across these downstream planes are discussed here. The peak probability for smaller droplet diameters decreases as the spray evolves from 15D to 45D. This trend reflects the breakup of larger droplets into smaller ones and the progressive redistribution of droplet sizes downstream. The cases exhibit a broader tail for droplet sizes beyond 100 µm, indicating that while many smaller droplets are formed, some large droplets persist in the downstream regions. However, the distributions become smoother with distance, especially by 45D, suggesting secondary breakup events stabilize over time. The inset plot with a zoomed section highlights the small probability values for larger droplets (above 100 µm). Even at longer distances, some large droplets survive, albeit with diminishing probability. In order to understand the contribution of larger droplets towards droplet size characteristics, the droplet PDP alone is not sufficient. A CDP based on droplet volume is therefore additionally analyzed for these cases.

For Case 1, a narrower distribution at smaller diameters (e.g., ~40 µm) for all planes is observed. The shift in peak probability from 15D to 45D suggests a gradual breakup due to lower jet velocity and ambient crossflow air speed. Droplet sizes are generally more tightly distributed with less secondary atomization, as indicated by the more defined single peak around 40-50 µm. In Case 1, stabilization occurs around ~45D, with fewer droplets and larger droplet sizes, and the distribution is broader, suggesting incomplete breakup/partial atomization. For Case 4, a broader distribution with a more pronounced bimodal shape (two peaks), especially for 15D and 30D. The first peak around 30-40 µm reflects smaller droplets formed early in the breakup process. A second peak at 75-100 µm suggests incomplete secondary atomization. Higher jet velocity leads to more energetic atomization, producing a wider range of droplet sizes. In Case 4, droplet sizes stabilize around 30D–35D, indicating that most of the breakup has already occurred by then. Case 11 ($V_{jet}$ = 30 m/s) achieves better atomization and stabilization with smaller, more uniform droplets across the downstream distances compared to Case 6 ($V_{jet}$ = 19 m/s). Both cases show peaks at 45–50 µm, but Case 6 has more variability and persistence of larger droplets, whereas Case 11 achieves tighter size control.

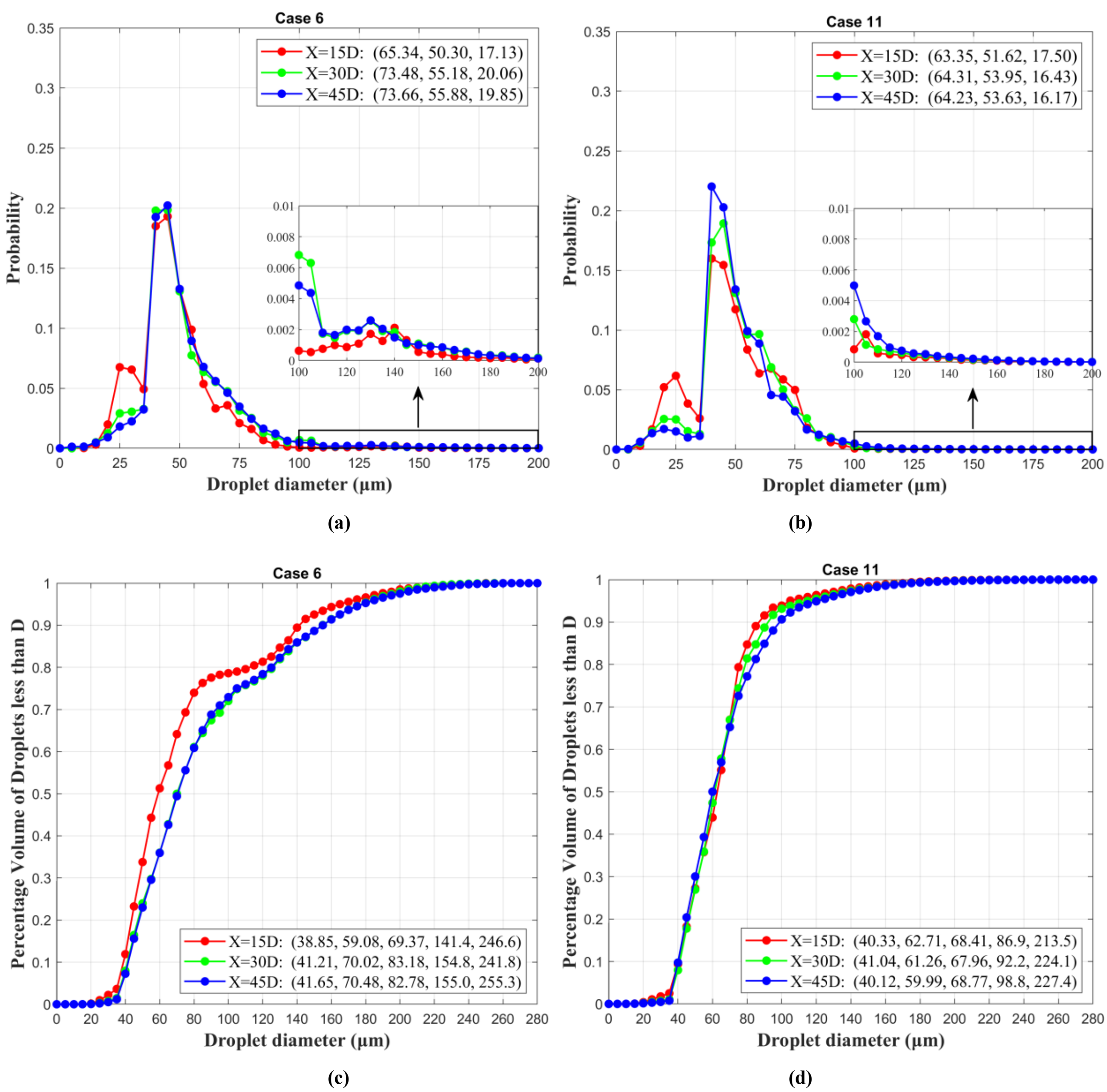

FIG 11. Droplet size distribution in the downstream: (a) Case 6-PDP (b) Case 11-PDP (c) Case 6-CDP (d) Case 11-CDP

Higher liquid jet velocity (as in Cases 4 and 11) is preferable for applications requiring finer spray distribution and efficient mixing due to the higher droplet count and smaller average size. Lower liquid jet velocities (Cases 1 and 6) might be better for coarser spray applications where moderate droplet sizes and less evaporation are desired. These results highlight the importance of jet velocity and breakup dynamics in controlling droplet size distributions and downstream behavior in such atomization processes.

### C.2 EFFECT OF PRESSURE

Figure 12 shows the effect of ambient pressure on the SMD and STD at constant liquid jet (19m/s) and crossflow velocities (61 & 65 m/s). Increasing crossflow pressure (2.1 bar to 3.8 bar) reduced the momentum flux ratio (q) (from 41 to 19), decreasing jet penetration, while the crossflow Weber number increases (from 71 to 150). Increasing ambient pressure increases the crossflow air density and drag force, promoting the breakup of larger droplets into smaller ones, which reduces the overall SMD. Since the SMD calculation is more sensitive to the presence of larger droplets, their reduction consequently lowers the SMD. A similar trend is observed in the STD, indicating that higher pressure promotes a more uniform droplet size distribution through enhanced atomization. However, it's important to note that the increased drag forces associated with higher pressure also lead to a reduction in the liquid jet penetration.

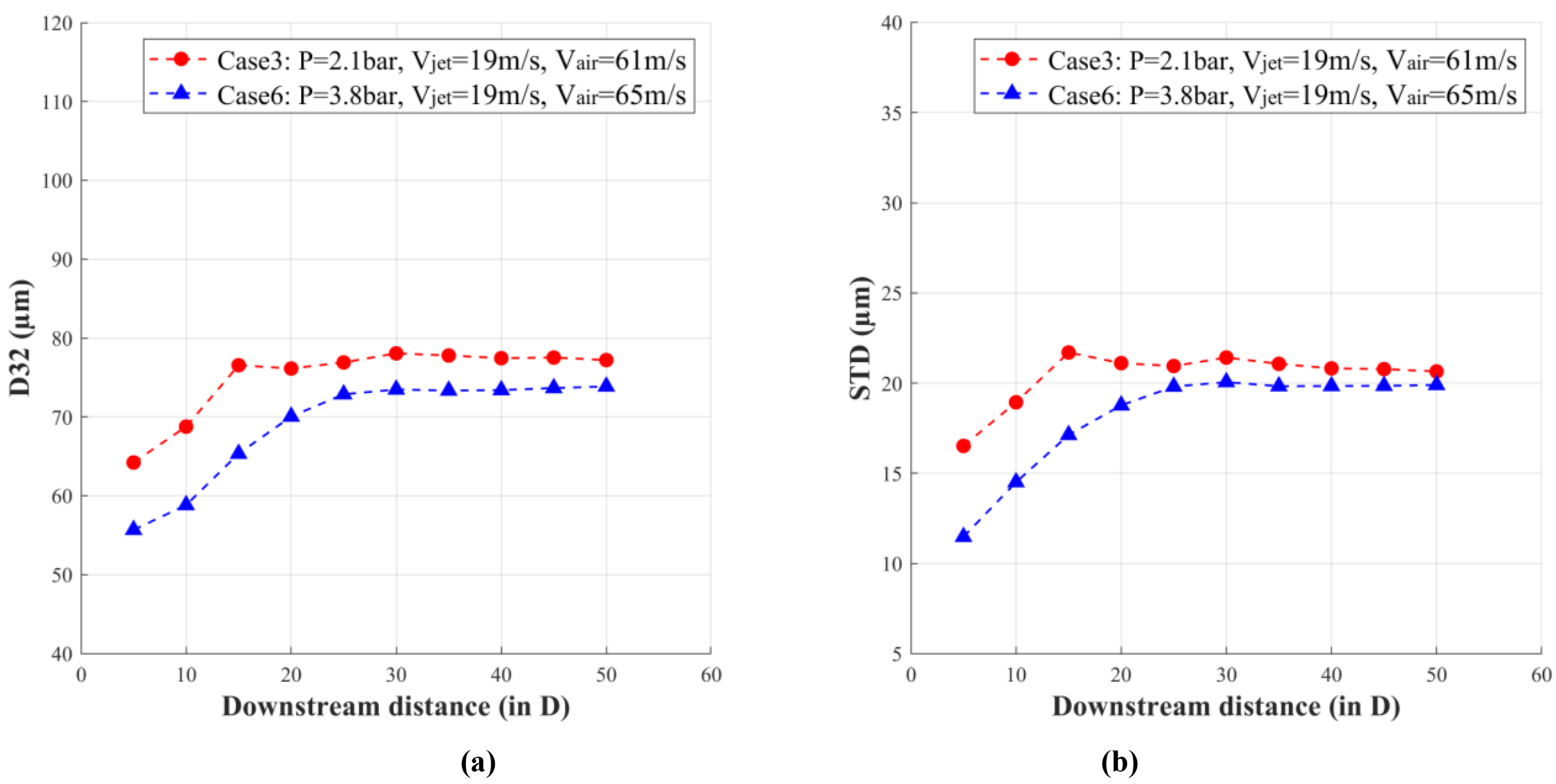

FIG 12. Effect of crossflow pressure on droplet size characteristics in the downstream: (a) D32, (b) STD

As discussed above, high crossflow pressure (Case 6) increases aerodynamic forces, enhancing primary breakup and producing smaller droplets. This is evident from the reduced breakup location, decreasing from $X_{breakup}$=9.64D in Case 3 to $X_{breakup}$=8.67D in Case 6. Lower crossflow pressure (Case 3) results in incomplete atomization, with larger droplets surviving further downstream due to weaker aerodynamic and shear forces. Higher pressure promotes the formation of smaller, stable droplets but reduces the overall droplet count. Case 3 shows greater penetration and broader spread, leading to more droplets in specific crossflow planes than Case 6. Additionally, the average droplet residence time in the domain is longer in Case 3 than in Case 6.

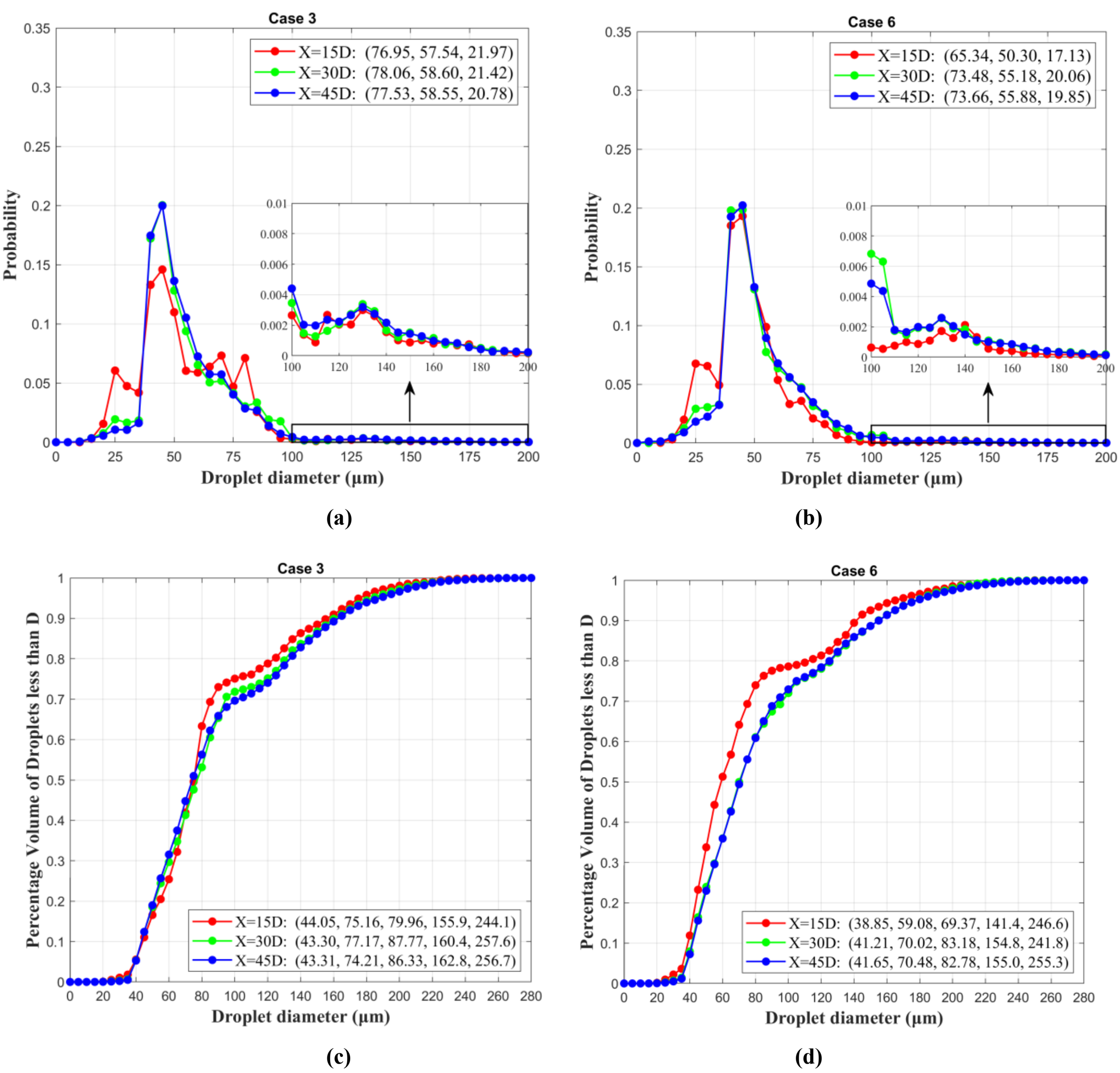

FIG 13. Droplet size distribution in the downstream: (a) Case 3-2.1bar (PDP) (b) Case 6-3.8bar (PDP) (c) Case 3-2.1bar (CDP) (d) Case 6-3.8bar (CDP)

In Fig. 13, the droplet size distributions for Case 3 (P = 2.1 bar) and Case 6 (P = 3.8 bar) at downstream distances of 15D, 30D, and 45D reveal distinct patterns influenced by crossflow pressure. In both cases, the primary peak occurs around 45–50 µm, but Case 6 shows sharper peaks with fewer large droplets, indicating more efficient atomization. The broader distribution in Case 3, with a secondary peak near 25 µm and a more prominent tail beyond 100 µm, suggests incomplete breakup and higher droplet variability. As droplets move downstream, both cases exhibit stabilization in their size distribution, but Case 6 maintains smaller droplets and tighter size control across all locations. This is reflected in the lower SMD and standard deviation values for Case 6, which are driven by stronger aerodynamic breakups due to higher ambient pressure. Case 6 is more suitable for high-performance spray systems like fuel injection, where smaller and more consistent droplet sizes are desired. Case 3 may be better suited for applications requiring larger droplets, such as agricultural spraying or fire suppression under ambient conditions, where variability in size can enhance coverage and droplet penetration. These results emphasize the importance of crossflow pressure in achieving the desired droplet size distribution and variability control in atomization processes.

### C.3 EFFECT OF CROSS FLOW VELOCITY

Figure 14 illustrates the influence of crossflow velocity on droplet size characteristics across various downstream locations. As expected, the results demonstrate a decrease in droplet size with increasing crossflow velocity. The interplay between aerodynamic drag and the momentum of the liquid jet can explain this trend. At lower crossflow velocities, the air exerts a weaker drag on the liquid jet, resulting in a higher momentum flux ratio (liquid momentum as compared to air momentum) and, hence, the deeper jet penetration. This deeper penetration leads to a broader spray plume but reduced atomization as larger droplets persist due to weaker shear forces. Consequently, the breakup is less effective, resulting in larger droplet sizes at lower crossflow velocities.

In contrast, at higher crossflow velocities, the drag force from the air increases significantly, bending and flattening the liquid jet into a flattened, sheet-like structure at the windward side. This reduced penetration flattens the trajectory of the liquid jet further, intensifying the effects of shear-driven K-H instabilities. This deformation corresponds to an increased air Weber number ($We>110$) and low momentum flux ratio, marking the dominance of aerodynamic shear over other forces and promoting a shift toward a "pure shear" breakup mode. The flattening jet triggers K-H instabilities at the liquid-air interface, which arise due to the velocity shear at the liquid-air interface. These instabilities amplify as surface waves on the jet's core grow in intensity, breaking the jet into ligaments and smaller droplets. The increased deformation leads to an earlier onset of primary breakup, transitioning the instability regime from Rayleigh-Taylor to Kelvin-Helmholtz instabilities as the crossflow velocity increases.[7] This stronger shear promotes extensive breakup and atomization into smaller droplets, reducing the overall $D_{32}$ and STD. In this mode, larger droplets disintegrate into smaller ones, further decreasing $D_{32}$ and leading to a finer and more uniform droplet size distribution.

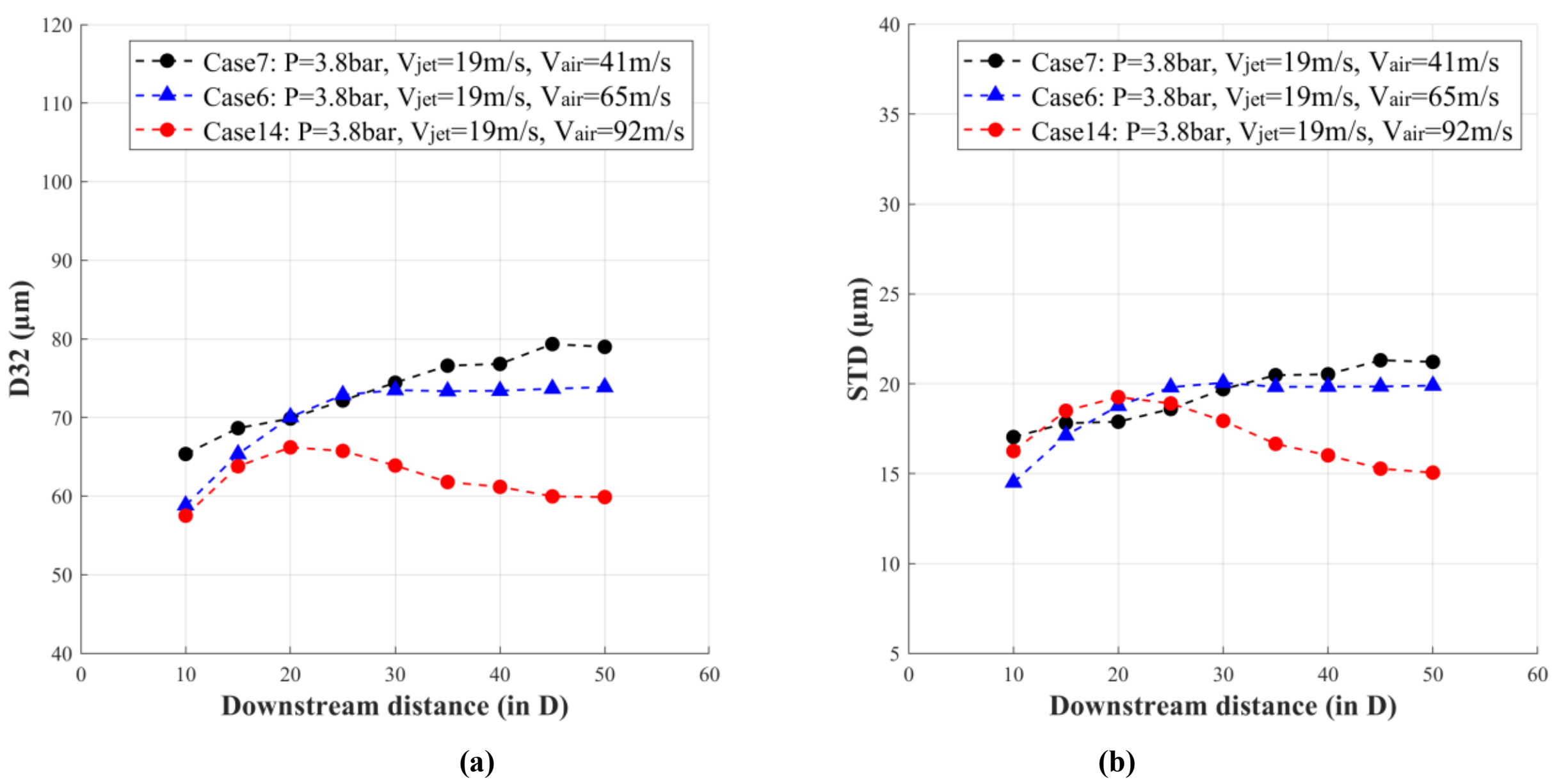

FIG 14. Effect of crossflow velocity on droplet size characteristics in the downstream: (a) $D_{32}$, (b) STD

As the crossflow velocity increases from 41m/s to 92m/s, the variation in $D_{32}$ changes from linear to a flattened bell curve. Higher crossflow velocities, as in Case 14, result in smaller droplet sizes across all downstream distances. At a higher crossflow velocity of 92m/s, $D_{32}$ stabilizes to around 60 µm beyond 35D, implying an effective atomization being attained. Lower crossflow velocities (Case 7) produce larger droplets, with $D_{32}$ exceeding 79 µm at 50D, indicating incomplete breakup and droplet coalescence downstream. Intermediate velocity (Case 6) leads to a moderate $D_{32}$ (72–74 µm), reflecting a balanced breakup process. Higher crossflow velocities (Case 14) exhibit the smallest variability, with STD decreasing to 15 µm around 45D. This reflects a more uniform droplet size distribution due to stronger aerodynamic shear forces. This promotes a stable atomization with minimal coalescence or growth downstream. Lower crossflow velocities (Case 7) result in weaker aerodynamic forces, allowing larger droplets to persist and coalesce, increasing size variability. This results in increased size variability (STD ~21 µm), and larger droplets surviving downstream indicate incomplete atomization. Case 6, with intermediate crossflow velocity, reflects a balanced atomization process, producing moderate droplet sizes and variability.

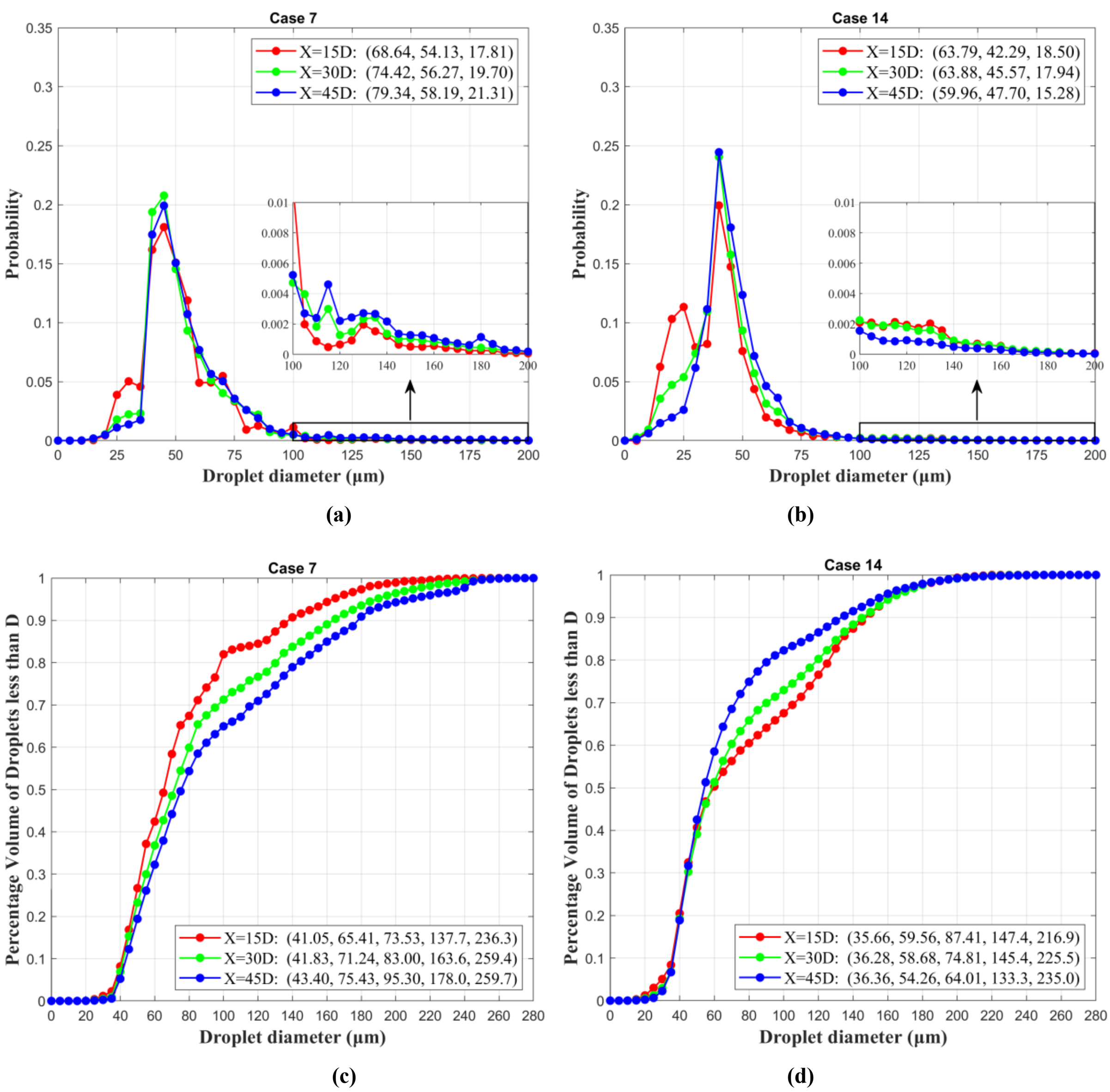

FIG 15. Droplet size distribution in the downstream: (a) Case 7: $V_{air}$=41m/s (PDP) (b) Case 14: $V_{air}$=92m/s (PDP) (c) Case 7: $V_{air}$=41m/s (CDP) (d) Case 14: $V_{air}$=92m/s (CDP)

The downstream droplet size distributions for Case 7 ($V_{air}$ = 41 m/s) and Case 14 ($V_{air}$ = 92 m/s) in Fig. 15 show the effect of crossflow velocity on atomization in the downstream. In Case 7, a noticeable peak of around 45 μm is present across all downstream planes, and with a secondary small peak near 25 μm in the 15D and 30D planes. This indicates an incomplete breakup, allowing larger droplets to persist. In the downstream evolution of droplet sizes, $D_{32}$ increases from 68.64 μm at 15D to 79.34 μm at 45D, and this can occur due to two factors: One from the conversion of droplets from Eulerian to lagrangian (droplet generation) and another due to droplet coalesce in the downstream (droplet growth). This is consistent with the lower crossflow velocity, where weaker aerodynamic forces allow droplet growth and size variability. The distribution becomes broader downstream, as indicated by the increasing $D_{32}$ and STD, suggesting a less uniform size distribution with larger droplet sizes.

Thus, lower crossflow velocity, as in Case 7, leads to less efficient breakup, resulting in larger droplets and a broader size distribution downstream.

Higher crossflow velocity, as in Case 14, results in finer, uniform droplets with a consistent peak of around 40 µm across the downstream planes, indicating more efficient atomization and a tighter droplet size distribution. $D_{32}$ stabilizes to around 60 µm, demonstrating that higher crossflow velocity (92 m/s) effectively limits droplet growth and coalescence. Droplets remain smaller and more uniform even at longer distances (45D). The distribution is narrower compared to Case 7, with fewer larger droplets present, reflecting enhanced breakup and better size control due to the stronger aerodynamic forces. These results suggest that higher crossflow velocities are ideal for applications requiring finer, more consistent droplet sizes. In comparison, lower velocities are suited for scenarios where larger droplets and broader distributions are acceptable.

### C.4 EFFECT OF CROSSFLOW WEBER NUMBER

The Weber number (We) represents the ratio of inertial forces to surface tension forces in the flow, and it plays a crucial role in droplet breakup. By keeping the momentum flux ratio constant (by adjusting both crossflow and liquid jet velocities proportionally), the results isolate the sole effect of the Weber number on droplet size characteristics.

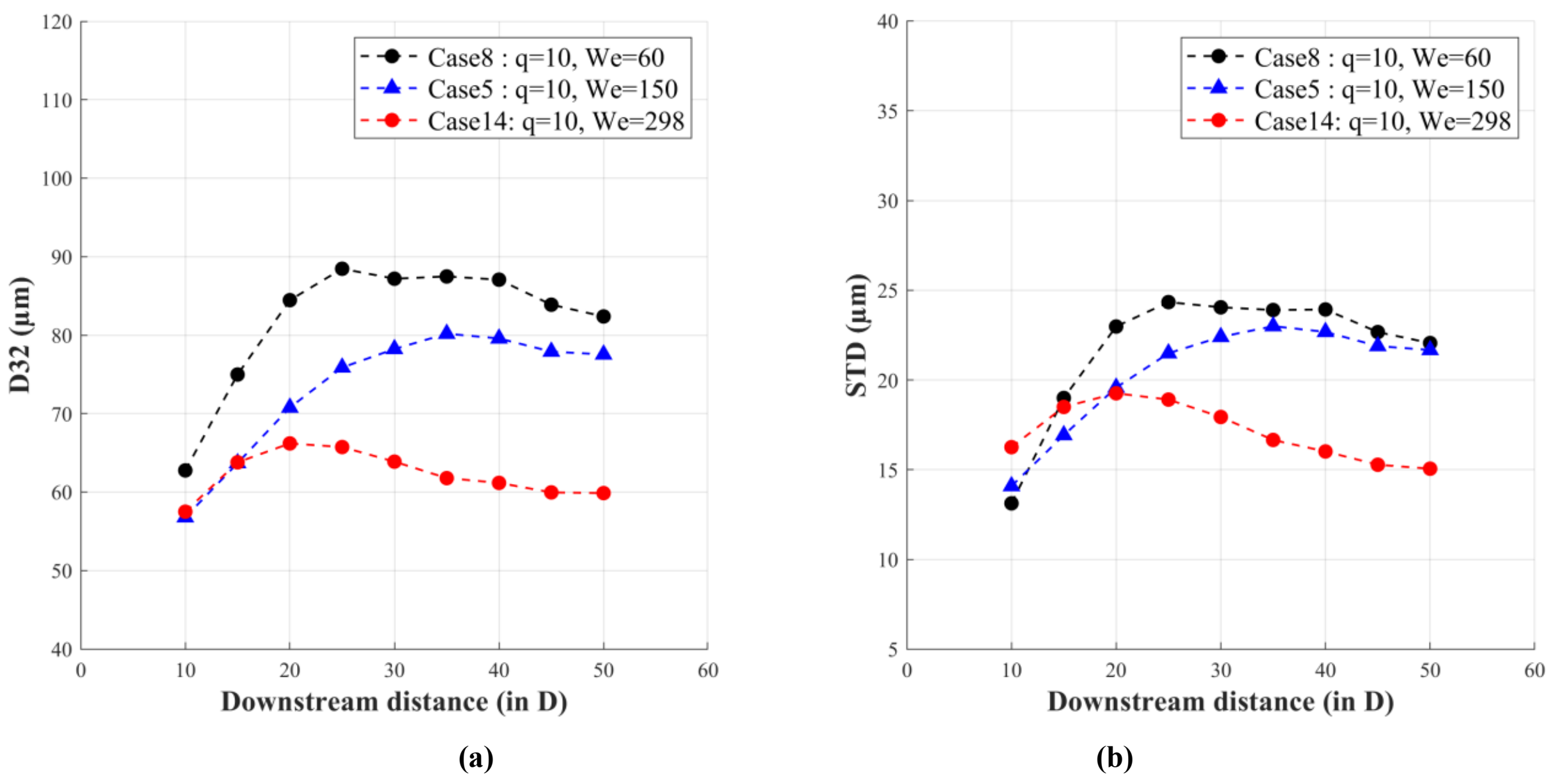

FIG 16. Effect of crossflow Weber number on droplet size characteristics in the downstream: (a) D32, (b) STD

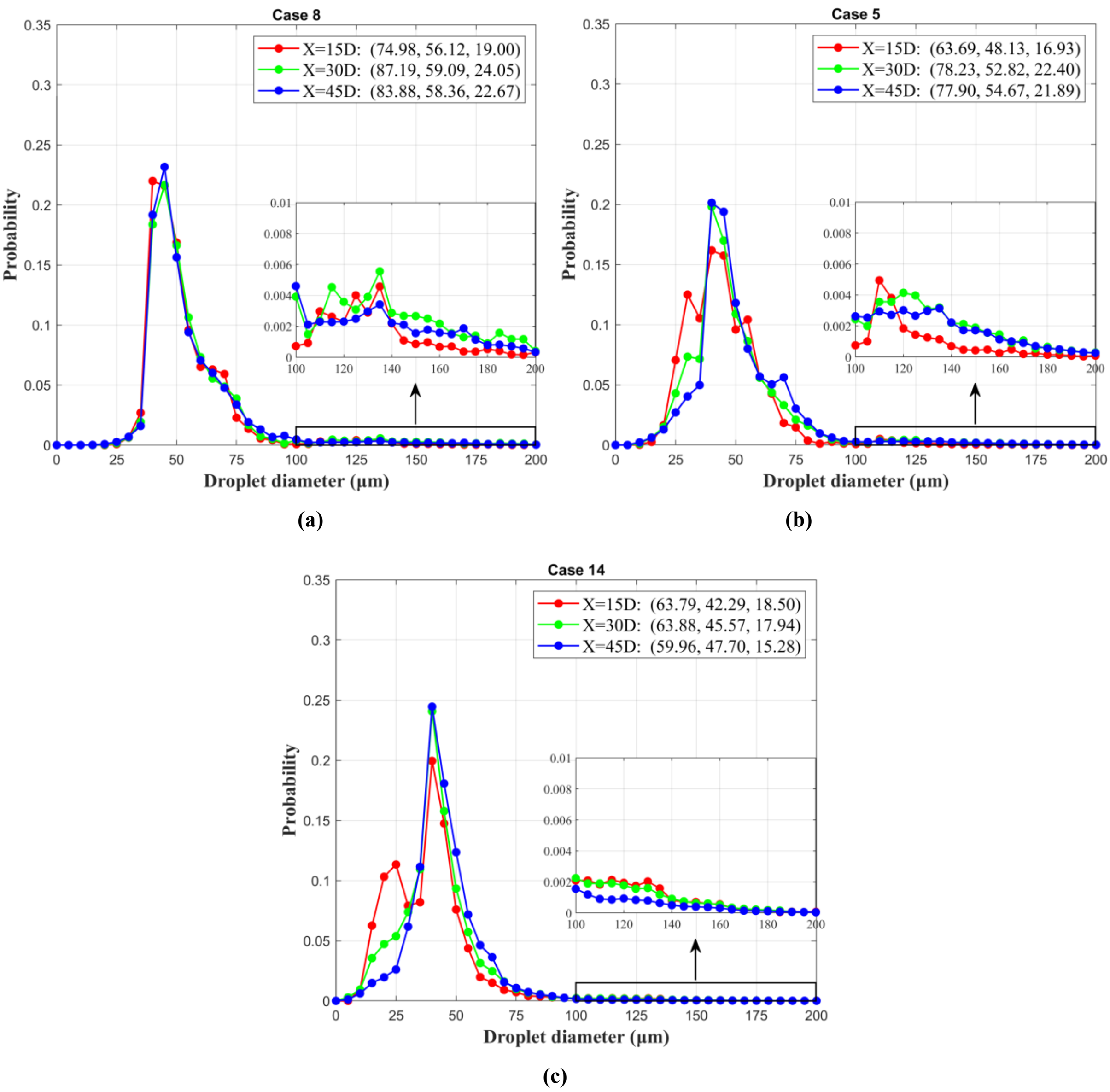

FIG 17. Droplet size distribution in the downstream: (a) Case8: We=60 (b) Case 5: We=150 (c) Case 14: We=298

Higher Weber numbers (Case 14, We = 298) result in smaller droplets, with $D_{32}$ stabilizing around 60 μm beyond 30D, indicating that stronger inertial forces dominate surface tension, enhancing atomization and promoting a uniform droplet size distribution with lower STD (~15 μm at 45D). In contrast, lower Weber numbers (Case 8, We = 60) show larger droplets (~80 μm) persisting downstream with higher STD (~25 μm at 45D), reflecting weaker breakup and greater size variability. Case 5 (We = 150) exhibits intermediate behavior, with $D_{32}$ around 70 μm and moderate STD (~20 μm), balancing atomization efficiency and droplet variability. Across all cases, increasing the Weber number enhances breakup, reducing droplet size and variability, as higher inertial forces overcome surface tension, resulting in finer, more consistent droplets. Similar aspects are already explained in the previous section.

### D. MEAN DROPLET SIZES AND DISPERSION CHARACTERISTICS

Table III. presents the statistical parameters of droplet size distributions across varying operating conditions in Liquid Jet in Crossflow (LJICF) systems. The cases span a range of momentum flux ratios (q), Weber numbers (We), and associated parameters to illustrate their effects on droplet characteristics. Mean droplet diameters ($D_{10}$, $D_{20}$, $D_{30}$, $D_{21}$, $D_{32}$) and representative diameters indicate finer droplets at higher q and We. Cumulative volume-based diameters ($D_{0.1}$, $D_{0.5}$, $D_{0.632}$, $D_{0.9}$, $D_{0.999}$) demonstrate shifts in droplet size distribution, emphasizing reduced dispersion (STD) at higher momentum flux ratios. Dispersion parameters, including the droplet uniformity index (DUI) and the MMD/SMD ratio, further quantify the effects of q and We. Typically, in the literature[2,3], the reported MMD/SMD ratio is 1.2, while in our simulations, the values observed are slightly above 1. The lower MMD/SMD ratio observed in the simulations compared to the typical 1.2 reported in literature likely stems from differences in atomization modeling, initial conditions, or numerical resolution, leading to narrower droplet size distributions. High Weber number cases (e.g., Cases 11, 12, 15, 16) exhibit lower DUI and MMD/SMD values, signifying more uniform droplet sizes with tighter distributions.

TABLE III. Statistical parameters for droplet size distribution at different operating conditions

| *Case No.* | *q* | *We* | *Mean diameters* | | | | | *Representative diameter* | | | | | *Dispersion parameters* | | | | |
|---|---|---|---|---|---|---|---|---|---|---|---|---|---|---|---|---|---|
| | | | *$D_{10}$ (μm)* | *$D_{20}$ (μm)* | *$D_{30}$ (μm)* | *$D_{21}$ (μm)* | *$D_{32}$ (μm)* | *$D_{0.1}$ (μm)* | *$D_{0.5}$ (μm)* | *$D_{0.632}$ (μm)* | *$D_{0.9}$ (μm)* | *$D_{0.999}$ (μm)* | *STD (μm)* | *MMD /SMD* | *DUI* | $\Delta_s$ | $\Delta_B$ |
| 1 | 8 | 71 | 58.67 | 62.85 | 68.85 | 67.33 | 82.64 | 48.37 | 84.75 | 121.86 | 189.1 | 266.7 | 22.54 | 1.026 | 0.568 | 1.76 | 2.28 |
| 2 | 16 | 71 | 57.88 | 61.62 | 66.93 | 65.61 | 78.97 | 47.53 | 78.96 | 105.68 | 175.2 | 265.3 | 21.14 | 1.000 | 0.552 | 1.72 | 2.51 |
| 3 | 41 | 71 | 57.48 | 61.07 | 65.74 | 64.88 | 76.18 | 47.68 | 77.31 | 89.42 | 164.2 | 260.9 | 20.61 | 1.015 | 0.474 | 1.61 | 2.52 |
| 4 | 66 | 71 | 57.52 | 60.60 | 64.28 | 63.84 | 72.34 | 47.55 | 74.37 | 84.24 | 136.3 | 255.0 | 19.06 | 1.028 | 0.377 | 1.28 | 2.60 |
| 5 | 10 | 150 | 53.15 | 57.24 | 62.81 | 61.64 | 75.65 | 45.52 | 76.95 | 107.40 | 172.7 | 260.9 | 21.24 | 1.017 | 0.560 | 1.76 | 2.58 |
| 6 | 19 | 150 | 54.98 | 58.30 | 62.52 | 61.82 | 71.90 | 46.13 | 73.65 | 84.96 | 154.5 | 257.9 | 19.38 | 1.024 | 0.452 | 1.57 | 2.68 |
| 7 | 48 | 60 | 56.46 | 59.82 | 64.43 | 63.38 | 74.75 | 47.15 | 75.42 | 87.84 | 170.9 | 263.3 | 19.77 | 1.009 | 0.513 | 1.75 | 2.67 |
| 8 | 10 | 60 | 58.08 | 62.16 | 68.40 | 66.52 | 82.84 | 48.20 | 81.95 | 130.50 | 199.0 | 268.0 | 22.14 | 0.989 | 0.645 | 1.94 | 2.38 |
| 9 | 77 | 60 | 55.78 | 58.44 | 61.76 | 61.23 | 68.96 | 46.38 | 70.24 | 78.82 | 135.4 | 258.2 | 17.42 | 1.018 | 0.390 | 1.36 | 2.86 |
| 10 | 77 | 38 | 60.14 | 63.83 | 68.68 | 67.75 | 79.52 | 48.91 | 81.74 | 96.08 | 167.4 | 267.7 | 21.39 | 1.028 | 0.461 | 1.54 | 2.41 |
| 11 | 48 | 150 | 53.84 | 56.51 | 59.27 | 59.32 | 65.18 | 45.71 | 68.33 | 75.32 | 100.3 | 225.1 | 17.17 | 1.048 | 0.291 | 0.86 | 2.48 |
| 12 | 77 | 150 | 53.00 | 56.01 | 59.05 | 59.20 | 65.62 | 45.63 | 68.65 | 75.94 | 106.0 | 222.3 | 18.11 | 1.046 | 0.308 | 0.95 | 2.43 |
| 13 | 19 | 60 | 57.98 | 61.58 | 66.77 | 65.39 | 78.52 | 47.48 | 79.37 | 97.49 | 181.0 | 269.0 | 20.73 | 1.011 | 0.551 | 1.79 | 2.53 |
| 14 | 10 | 298 | 46.24 | 49.19 | 52.95 | 52.31 | 61.39 | 41.96 | 59.75 | 71.72 | 141.8 | 236.1 | 16.67 | 0.973 | 0.509 | 1.83 | 3.23 |
| 15 | 20 | 298 | 51.83 | 54.17 | 56.61 | 56.62 | 61.85 | 44.52 | 63.58 | 70.71 | 93.2 | 216.3 | 15.77 | 1.028 | 0.302 | 0.84 | 2.61 |
| 16 | 49 | 298 | 46.99 | 50.48 | 53.61 | 54.38 | 60.48 | 44.14 | 63.04 | 69.37 | 91.1 | 201.1 | 18.70 | 1.042 | 0.281 | 0.82 | 2.37 |

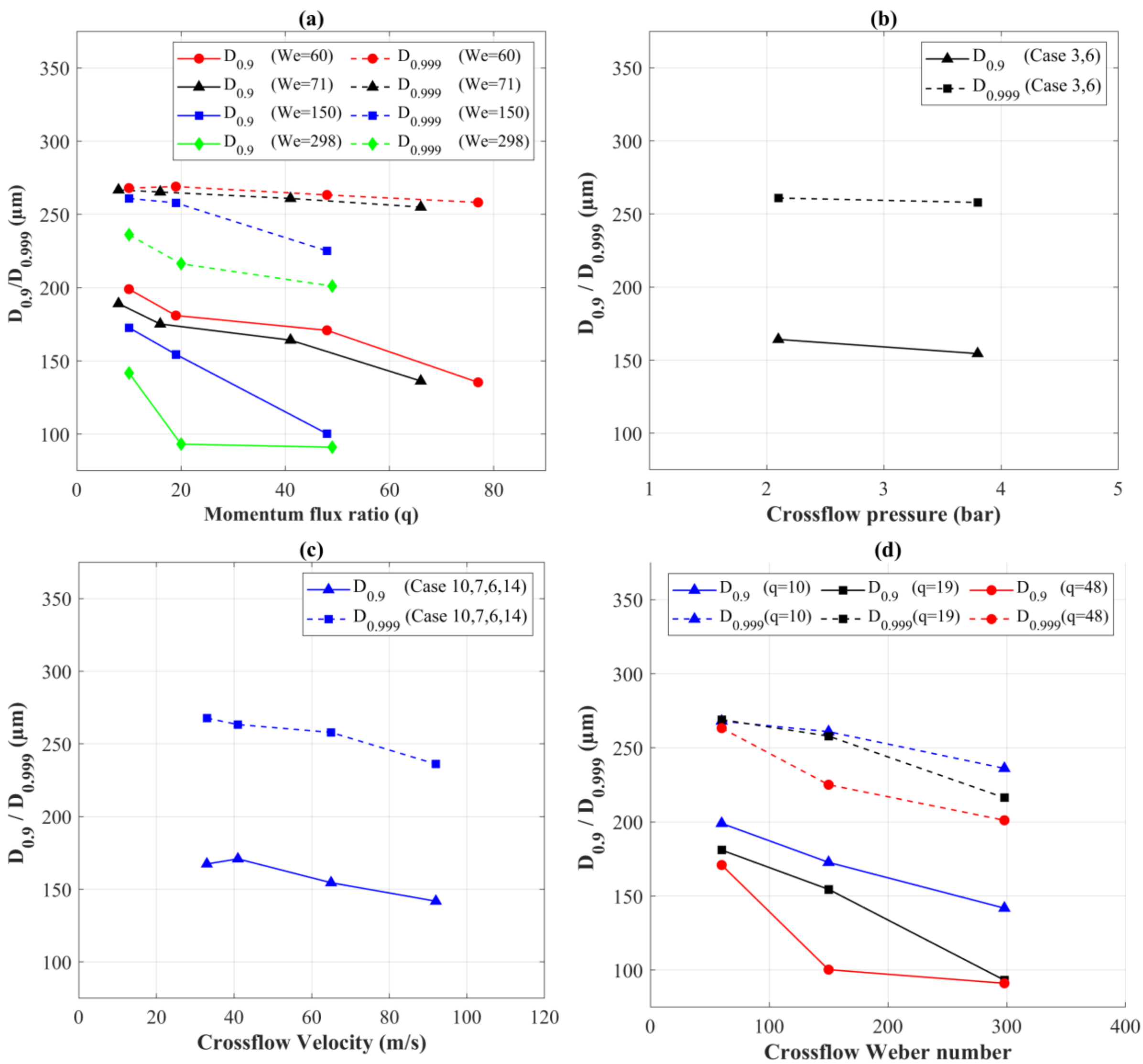

FIG 18. Maximum droplet sizes variation with (a) Momentum flux ratio (b) Crossflow pressure (c) Crossflow velocity (d) Crossflow Weber number

At low Weber numbers (We=60 and We=71), the variation in $D_{0.999}$ with increasing momentum flux ratio (q) is minimal. However, at higher Weber numbers (We=150 and We=298), the variation in $D_{0.999}$ becomes more pronounced, as observed in Fig. 18 (a). At low Weber numbers, the aerodynamic forces acting on the liquid jet are relatively weaker than surface tension forces. Consequently, $D_{0.999}$ does not vary significantly with q, as the breakup mechanism is dominated by inertial and surface tension effects rather than aerodynamic forces. At high Weber numbers, the aerodynamic forces become much stronger relative to surface tension forces, enhancing the atomization process. Under these conditions, the variation in q significantly influences the breakup process, leading to a noticeable reduction in $D_{0.999}$ as q increases. This is because the aerodynamic forces at high Weber numbers are more effective in fragmenting the larger droplets, which $D_{0.999}$ represents. Thus, the variation in maximum droplet diameter ($D_{0.999}$) is highly dependent on Weber number, with significant variation occurring only at higher Weber numbers. Hence, it is observed that the variation in $D_{0.999}$ is less sensitive to the changes in q at lower Weber numbers. The narrowing gap between $D_{0.9}$ and $D_{0.999}$ with increasing q and We suggests that high-energy conditions promote uniform atomization and suppress the formation of outlier droplets.

As observed from Fig. 18 (b), the maximum droplet size ($D_{0.999}$) is relatively insensitive to crossflow pressure changes within the range studied (2.1 bar to 3.8 bar). Higher crossflow pressure enhances droplet breakup for the majority of the droplet population, resulting in a smaller $D_{0.9}$. This indicates that crossflow pressure has a more pronounced effect on the intermediate and smaller droplets than on the largest droplets.

Figure 18 (c) demonstrates the effect of crossflow velocity ($V_{air}$) on $D_{0.999}$ (maximum droplet diameter) and $D_{0.9}$ (droplet diameter encompassing 90% of the total droplet volume) across four cases with varying crossflow velocities. As crossflow velocity increases from 33 m/s (Case 10) to 92 m/s (Case 14), $D_{0.999}$ shows a gradual decrease. The decrease is more pronounced at higher velocities (from Case 6 to Case 14), indicating stronger aerodynamic effects at higher crossflow velocity. $D_{0.9}$ consistently decreases with increasing crossflow velocity, showing a similar reduction as compared to $D_{0.999}$. The trend highlights a progressive reduction in the size of the majority of the droplets as the crossflow velocity increases. Larger droplets (represented by $D_{0.999}$ are less affected by aerodynamic forces at lower crossflow velocities because of their high inertia. However, as the crossflow velocity increases, the relative velocity difference between the jet and the crossflow grows, leading to enhanced shear forces and partial breakup of these large droplets. The observed sharp decrease in $D_{0.9}$ and $D_{0.999}$ highlights the dominant role of crossflow velocity in influencing the size of the bulk droplet population.

Figure 18 (d) shows the effect of crossflow Weber number ($We=\rho_{air}V_{air}^2D_{jet}/\sigma$) on $D_{0.999}$ (maximum droplet diameter) and $D_{0.9}$ (droplet diameter encompassing 90% of the droplet volume) for different momentum flux ratios (q). While crossflow velocity and Weber number effects show similar outcomes, the Weber number consolidates the combined effects of crossflow velocity, fluid density, and surface tension into a single parameter. This makes it universally applicable for comparing atomization behavior across different fluids and operating conditions. $D_{0.999}$ decreases with increasing Weber number for all values of q. The rate of decrease is more significant at higher Weber numbers, indicating enhanced droplet breakup in regimes dominated by higher aerodynamic forces. Higher Weber numbers correspond to stronger aerodynamic forces (via the term $\rho_{air}V_{air}^2$), which enhance the breakup of droplets and lead to smaller $D_{0.999}$ and $D_{0.9}$. At higher Weber numbers, the breakup mechanism transitions to a more vigorous regime, leading to smaller maximum droplet sizes. $D_{0.9}$ follows a similar decreasing trend as $D_{0.999}$, but the reduction is more pronounced, especially at lower Weber numbers.

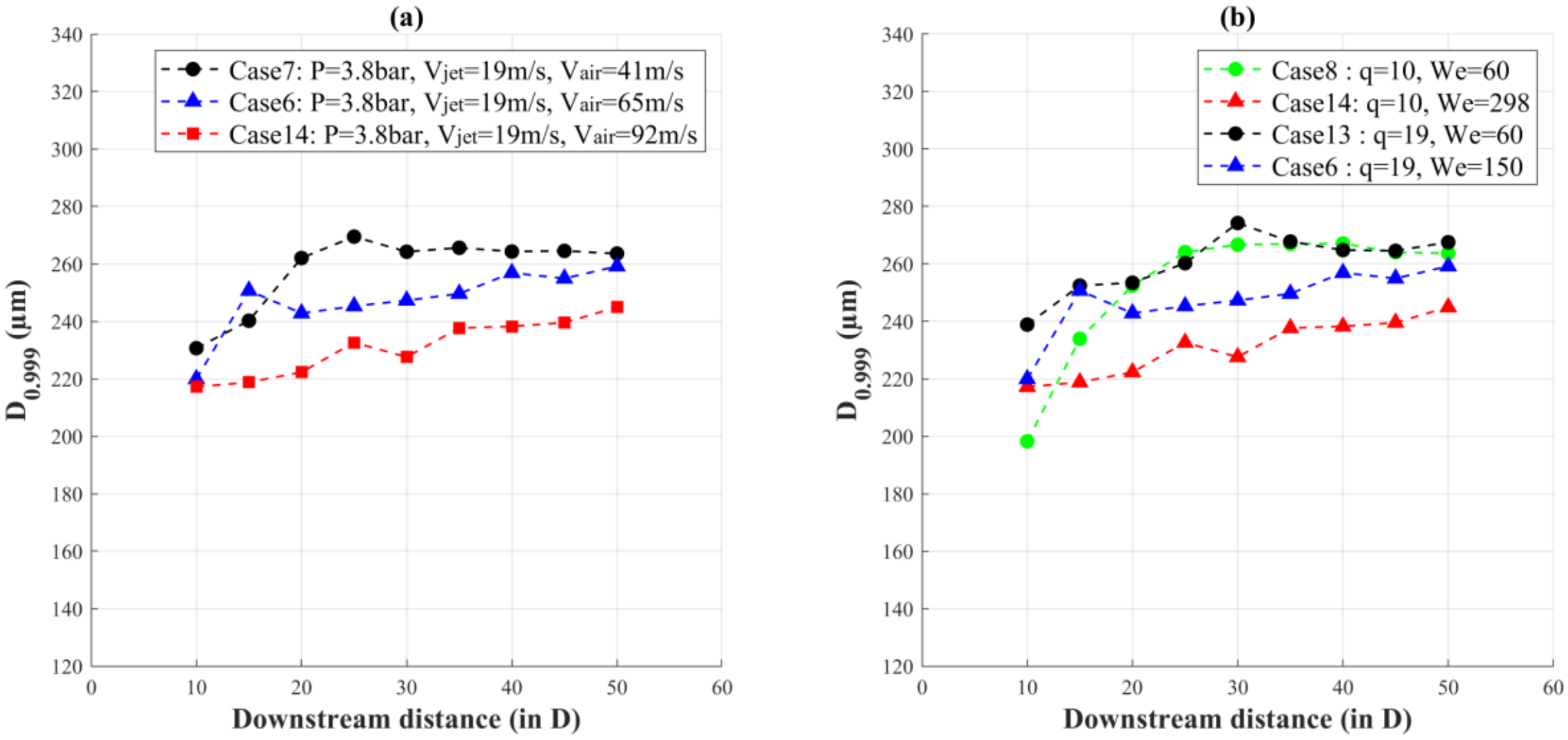

FIG 19. Downstream variation in maximum droplet sizes with (a) crossflow velocity and (b) crossflow Weber number and momentum flux ratio

Figure 19 illustrates the downstream variation of $D_{0.999}$ for different crossflow velocities, Weber numbers, and momentum flux ratios, highlighting the significant influence of crossflow velocity and Weber number on the atomization process. For Cases 8 and 13 (We=60), increasing q from 10 to 19 results in larger droplets (as a consequence of higher mass flow rate for q=19) in the 10–20D range, while the variation becomes similar beyond 20D. This supports the observation that $D_{0.999}$ is less sensitive to changes in q at lower Weber numbers, where aerodynamic forces are weaker. However, at higher Weber numbers (We=150, 298), $D_{0.999}$ shows a more pronounced variation with q, reflecting the dominance of aerodynamic forces over surface tension. The narrowing gap between larger droplets ($D_{0.999}$) and the bulk population ($D_{0.9}$) at higher energy conditions indicates a transition to a more uniform atomization regime. These results underline the importance of operating at higher Weber numbers and crossflow velocities to achieve efficient atomization and reduce the occurrence of larger, outlier droplets in crossflow spray systems.

## D.1 EFFECT OF VARIOUS PARAMETERS ON NUMBER DENSITY OF DROPLETS

This section explores the variations in the number density of droplets along the downstream distance with liquid jet and crossflow parameters. Higher velocities and pressures generally promote finer atomization and a larger number of smaller droplets, while lower velocities and pressures may lead to larger droplets and potentially fewer in number. The momentum flux ratio, which measures the relative strength of the jet to the crossflow, plays a critical role in droplet number density. A higher momentum flux ratio correlates with deeper jet penetration into the crossflow, expanding the spray plume and resulting in a greater number of droplets. Droplet residence time also influences the number of droplets observed. A longer residence time allows droplets to remain in the domain for a more extended period, contributing to the total droplet count. Conversely, shorter residence times may reduce the measured number of droplets due to faster dispersion and removal from the spray zone. Understanding these relationships is crucial for optimizing atomization processes to achieve desired droplet characteristics tailored to specific applications in industries such as aerospace, automotive, power generation, and more.

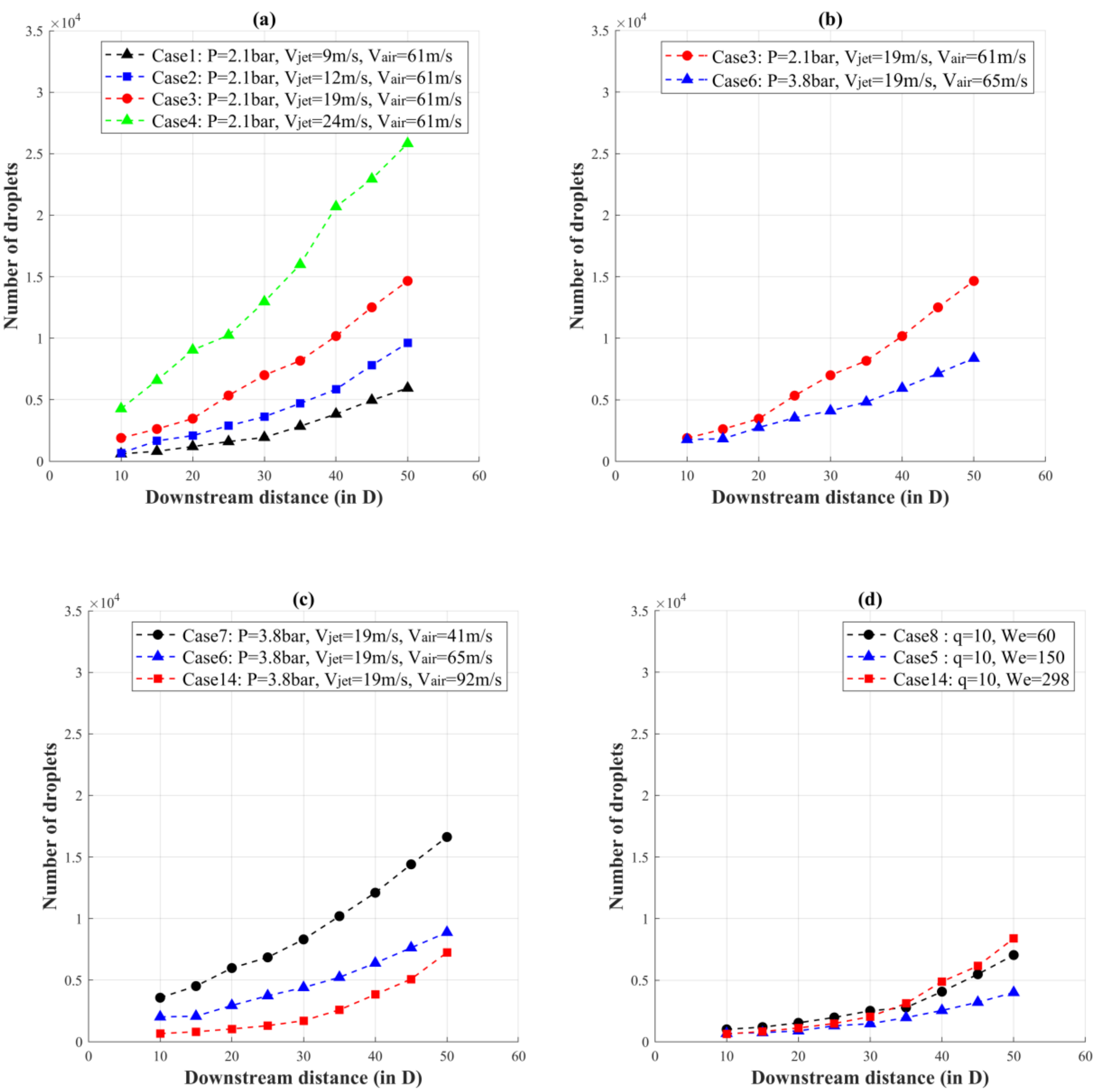

FIG 20. Variation in number density of droplets with downstream : (a) Effect of liquid jet velocity/momentum flux ratio (b) Effect of crossflow pressure (c) Effect of crossflow velocity (d) Effect of crossflow Weber number

Figure 20 (a) illustrates the effect of the liquid jet velocity/momentum flux ratio on the number density of droplets downstream. Higher jet velocities (e.g., Case 4: $V_{jet} = 24$ m/s) increase the momentum flux ratio (q), enabling deeper jet penetration and a broader spray plume. The increased mass flow rate at higher velocities contributes to a greater number of droplets per unit volume downstream. In contrast, lower jet velocities (Case 1:

$V_{jet}$ = 9 m/s) generate fewer droplets due to limited penetration and lesser mass flow rate, resulting in reduced interaction with the crossflow air, resulting in weaker breakup. Figure 20 (b) shows that higher crossflow pressure promotes finer droplets but reduces droplet number density in the 50D downstream region. This reduction can be attributed to the more confined spray plume caused by decreased jet penetration. Additionally, the increased aerodynamic drag at higher pressures deforms Eulerian droplets, potentially delaying their conversion to Lagrangian particles due to the conversion criteria (sphericity and minimum diameter).

Figure 21 shows that Case 6 retains more liquid mass in the Eulerian form compared to Case 3, with more elongated ligaments observed at higher pressure. This difference arises from the instability mechanisms: Case 3 exhibits Rayleigh-Taylor (R-T) instability, leading to quicker atomization, while Case 6 exhibits Kelvin-Helmholtz (K-H) instability with longer wavelengths, resulting in delayed atomization. At higher pressures, the average droplet velocity increases, accelerating their downstream transport. Consequently, the droplets spend less time within the domain, leading to a reduction in droplet residence time. These combined effects of confined plume, delayed particle conversion, and reduced residence time explain the observed decrease in droplet density at higher crossflow pressures.

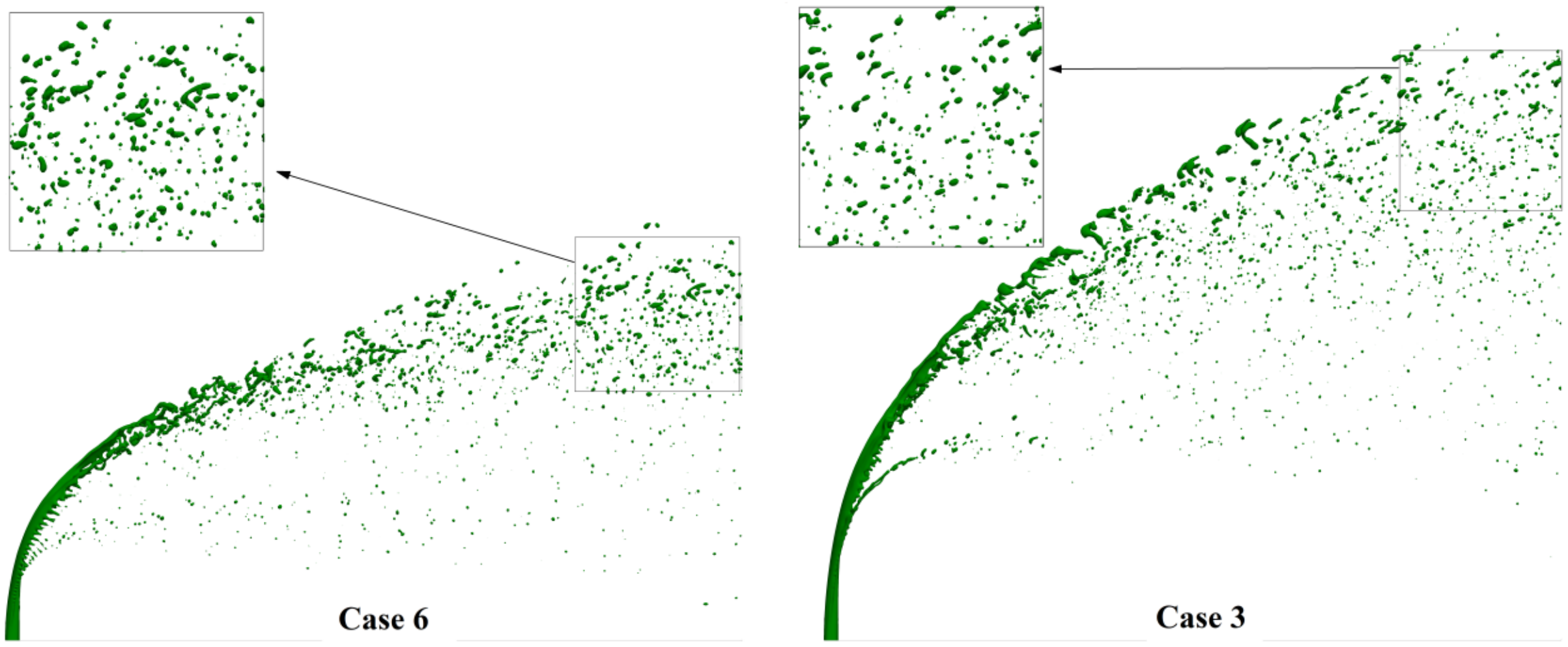

FIG 21. Eulerian liquid distribution shows highlighting ligaments and elongated droplets for Case 6 (P=3.8bar) and Case 3 (P=2.1bar). Higher number density in Case 6 indicates delayed particle conversion (Eulerian to Lagrangian) at increased pressure.

Figure 20 (c) shows that increasing crossflow velocity (Case 14: $V_{air}$ = 92 m/s) intensifies aerodynamic forces, decreasing the momentum flux ratio ($q= \rho_{jet}V_{jet}^2/\rho_{air}V_{air}^2$) of them but enhancing atomization and generating more droplets with a smaller size. Lower crossflow velocities (Case 7: $V_{air}$ = 41 m/s) lead to fewer droplets due to weaker breakup and reduced drag. However, as the crossflow velocity increases, the residence time of the droplet in the domain decreases. This, in turn, results in a lower droplet density in the downstream planes.

Weber number quantifies the relative influence of inertia and surface tension forces, determining the breakup regime and fragmentation behavior independent of velocity alone. However, a clear dependence on the Weber number is not observed in the droplet density in the downstream planes. Figure 20 (d) shows that high Weber numbers (We = 298, Case 14) result in more efficient atomization, with the generation of a large number of smaller droplets. Though the droplet residence time decreases with an increase in Weber number (from the increased crossflow velocity), the generation of more number of smaller droplets and the increased mass flow rate of the liquid jet (Case 8: $V_{jet}$=9m/s, Case 5: $V_{jet}$=14m/s, Case 14: $V_{jet}$=19m/s) are the observed causes for the small difference in the number of droplets.

In summary, higher jet velocities enhance penetration and spread, increasing droplet generation, while greater crossflow velocity and crossflow pressures promote finer breakup and droplet formation. The interplay between momentum flux ratio, spray plume size, and residence time governs droplet density along the downstream distance. Achieving optimal atomization and droplet distribution requires balancing jet velocity with crossflow parameters, as both fluid dynamics and jet-crossflow interactions play a crucial role in droplet formation.

### E. DROPLET SIZE DISTRIBUTION FUNCTIONS

The simulated droplet-size spectra for Cases 6 and 11 (Fig. 22a–d) have been fitted using Normal, Log-Normal, Rosin–Rammler (RR), and Modified Rosin–Rammler (MRR) distributions. As shown qualitatively in Fig. 22, the Log-Normal curve more faithfully reproduces the skewed tails of the distribution than the symmetric Normal model, while RR and MRR capture the cumulative volume trends over the full size range. To provide a quantitative comparison, we have evaluated the coefficient of determination ($R^2$) and the root-mean-square error (RMSE) for all four fits across Cases 1–16 (Table IV).

The results presented in Fig. 22 (a) and (b) indicate the effectiveness of these distributions in capturing the underlying droplet size trends. However, the Normal distribution captures the central tendency of the drop size data well. However, it tends to overpredict the probability density for intermediate droplet diameters (50-100μm) and underpredicts at the tails. This behavior is expected due to the symmetric nature of the Normal distribution, which does not account for the skewness observed in many spray atomization datasets. Across Cases 1–15, the log-normal distribution consistently outperforms the normal fit, exhibiting higher $R^2$ (0.737–0.867 vs. 0.496–0.734) and lower RMSE (0.0146–0.0231 vs. 0.0242–0.0319). This indicates that the droplet-size spectra in these conditions are inherently skewed, with a pronounced tail of small droplets that the log-normal model captures more accurately, indicating its suitability for extrapolating droplet sizes outside the measured range. The Log-Normal distribution provides a satisfactory fit to the measured data, allows extrapolation to larger droplet sizes, and aligns with the general characteristics of atomization, where smaller droplets dominate. It accurately captures the distribution's skewness and tail behavior and emphasizes the smaller droplet population while fitting intermediate sizes effectively. However, the sharp peak obtained in the simulation is not captured adequately. This sharp peak is partly the result of the CCL droplet conversion criteria employed in the simulation. Case 16 is the lone exception: here, the normal fit yields a superior $R^2$ (0.734 vs. 0.666) and lower RMSE (0.0191 vs. 0.0214), suggesting a more symmetric size distribution under those particular momentum-flux and Weber-number conditions. Overall, the log-normal form is the preferred general model for our LJICF spray data, with the possibility that certain high-q, high-We cases revert to near-Gaussian behaviour and may warrant a normal fit.

TABLE IV. Goodness of fit metrics (RMSE and $R^2$) for Normal, Log-normal, RR, Modified RR droplet size distribution fits (Cases 1–16).

| **Case** | **Normal $R^2$** | **Normal RMSE** | **Log-normal $R^2$** | **Log-normal RMSE** | **Rosin-Rammler $R^2$** | **Rosin-Rammler RMSE** | **Modified Rosin-Rammler $R^2$** | **Modified Rosin Rammler RMSE** |
|---|---|---|---|---|---|---|---|---|
| 1 | 0.5173 | 0.0294 | 0.7688 | 0.02035 | 0.9865 | 0.04012 | 0.9844 | 0.04308 |
| 2 | 0.5244 | 0.02991 | 0.7735 | 0.02064 | 0.9852 | 0.04247 | 0.9830 | 0.04544 |
| 3 | 0.5898 | 0.02571 | 0.8261 | 0.01674 | 0.9852 | 0.04300 | 0.9826 | 0.04655 |
| 4 | 0.6237 | 0.02415 | 0.8618 | 0.01464 | 0.9906 | 0.03502 | 0.9891 | 0.03769 |
| 5 | 0.6226 | 0.02437 | 0.8596 | 0.01486 | 0.9863 | 0.04019 | 0.9842 | 0.04318 |
| 6 | 0.6033 | 0.02563 | 0.8423 | 0.01616 | 0.9854 | 0.04239 | 0.9829 | 0.04585 |
| 7 | 0.5884 | 0.02670 | 0.8347 | 0.01691 | 0.9817 | 0.04683 | 0.9786 | 0.05070 |
| 8 | 0.4961 | 0.03198 | 0.7371 | 0.02310 | 0.9807 | 0.04698 | 0.9789 | 0.04914 |
| 9 | 0.6092 | 0.02633 | 0.8550 | 0.01604 | 0.9860 | 0.04193 | 0.9842 | 0.04447 |
| 10 | 0.5365 | 0.02751 | 0.7924 | 0.01841 | 0.9884 | 0.03830 | 0.9861 | 0.04203 |
| 11 | 0.6675 | 0.02293 | 0.8619 | 0.01477 | 0.9966 | 0.02114 | 0.9961 | 0.02260 |
| 12 | 0.6631 | 0.02299 | 0.8217 | 0.01673 | 0.9957 | 0.02362 | 0.9950 | 0.02538 |
| 13 | 0.5106 | 0.03091 | 0.7631 | 0.02150 | 0.9835 | 0.04432 | 0.9807 | 0.04789 |
| 14 | 0.6893 | 0.02380 | 0.8675 | 0.01554 | 0.9784 | 0.04871 | 0.9753 | 0.05211 |
| 15 | 0.6403 | 0.02591 | 0.8403 | 0.01727 | 0.9943 | 0.02663 | 0.9938 | 0.02785 |
| 16 | 0.7340 | 0.01907 | 0.6658 | 0.02137 | 0.9970 | 0.01937 | 0.9966 | 0.02070 |

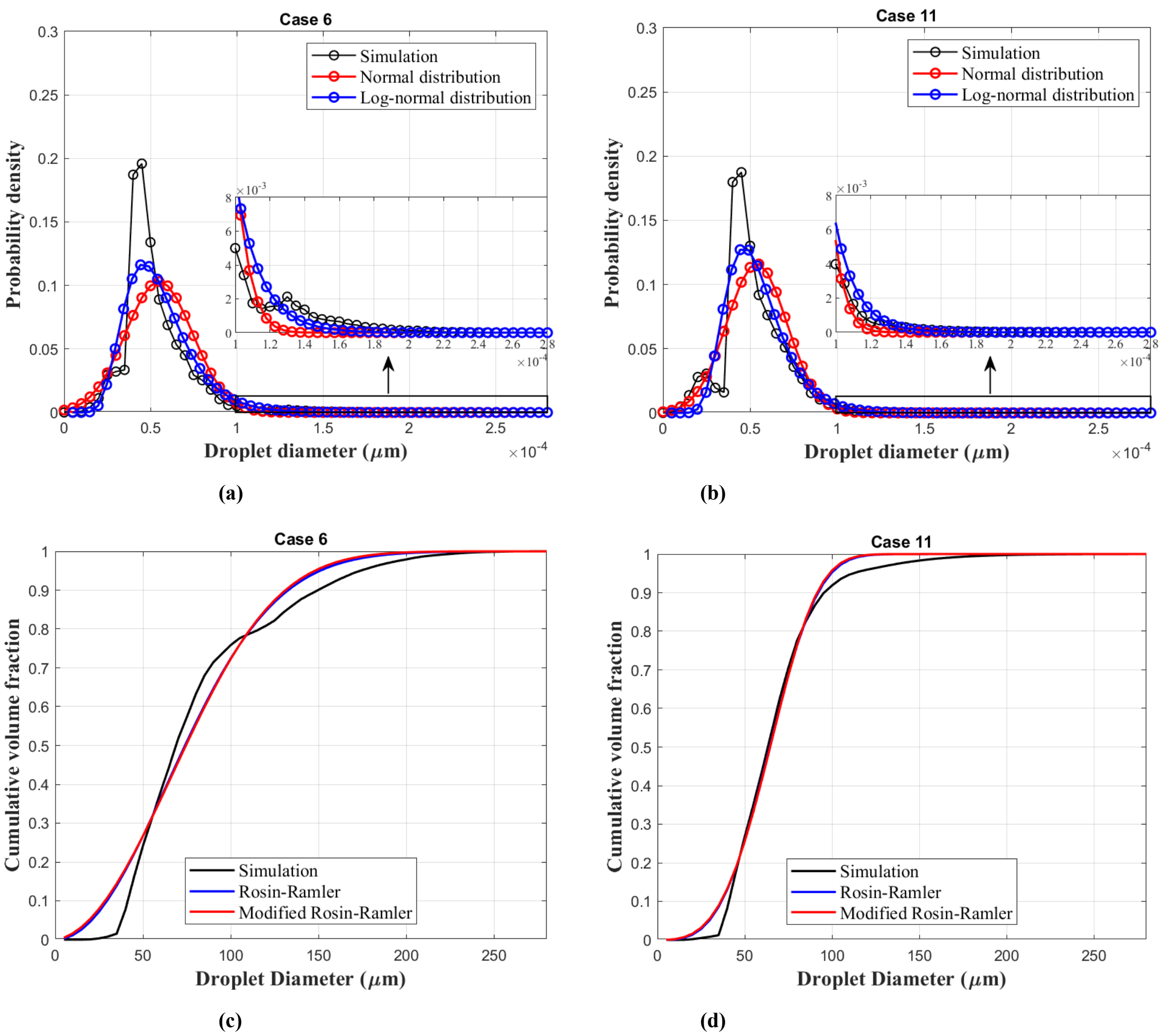

FIG 22. Comparison of measured droplet size distributions with Normal, Log-Normal, and Rosin-Ramler fits for Cases 6 and 11.

The Rosin-Ramler distribution is a widely used empirical model for characterizing droplet size distributions in spray atomization, using a two-parameter model (scale and shape parameters) to represent cumulative volume fractions effectively. In Cases 6 ($R^2 = 0.9854$, RMSE = 0.0424) and 11 ($R^2 = 0.9966$, RMSE = 0.0211), it captures the overall trend particularly well for intermediate and large droplet sizes, as evident from its close alignment with the simulation data. Across all 16 cases, the classical Rosin–Rammler fit achieves $R^2$ values from 0.981 to 0.997 and RMSE between 0.019 and 0.049, whereas the Modified Rosin–Rammler yields slightly lower $R^2$ (0.978–0.996) and marginally higher RMSE (0.021–0.052). While both forms far outperform the normal and log-normal models in cumulative-volume correlation, their point-wise PDF errors are roughly 1.5–2× larger than the log-normal RMSE (0.0146–0.0231), reflecting minor local deviations around sharp peaks and tails.

While the standard Rosin–Rammler model may underpredict the smallest droplets due to limited skewness control, it still provides sufficient flexibility for these datasets, and the modified RR model introduces no appreciable improvement with additional parameters or modification in the exponents. This absence of difference is likely to stem from: (1) the minimal skewness in the measured droplet size spectra, which already is fitting well within the standard Rosin-Ramler assumptions, (2) the standard two-parameter model is adequate to capture the dominant atomization physics here, rendering additional modifications unnecessary. In summary, the classical Rosin–Rammler distribution's inherent suitability and superior $R^2$/RMSE performance render the modified version unnecessary for our LJICF spray data. For future work, alternative models such as Nukiyama–Tanasawa or upper-limit functions could be explored to refine tail-behavior predictions further and deepen our understanding of the atomization process.

## V. CONCLUSION

This study presents a detailed analysis of droplet size distribution (DSD) in Liquid Jet in Crossflow (LJICF) systems, employing a robust VOF-LPT coupled framework. The computational model is rigorously validated against experimental data, demonstrating an average error below 10% for SMD predictions, underscoring its predictive reliability in spray atomization studies. The results reveal the significant impact of momentum flux ratio, Weber number, crossflow pressure, and velocity on atomization efficiency ($D_{32}$), droplet distribution (PDPs and CDPs), and downstream behavior. Key outcomes include:

- Larger droplets significantly influence droplet size characteristics, particularly the Sauter mean diameter ($D_{32}$), as observed from the CDPs and PDPs. In high-momentum flux cases (Cases 11 and 12), jet penetration and impingement influence droplet breakup. However, computational predictions show limited sensitivity to further velocity increases due to the lower cutoff in the VOF-LPT coupling. Higher velocities reduce droplet size standard deviation, indicating a more uniform DSD. Increased momentum flux ratios produce finer, more uniform droplets, as evident from steeper CDP curves and reduced STD.
- Higher crossflow pressures and velocities amplify aerodynamic forces, shifting the DSD towards smaller diameters and promoting efficient atomization, which is crucial for fuel-air mixing and related applications. This induces Kelvin-Helmholtz instabilities, promoting enhanced primary breakup and transitioning from Rayleigh-Taylor to shear-dominated breakup, resulting in narrower size distributions and smaller mean droplet diameters ($D_{32}$).

- Droplet sizes stabilize downstream, with higher jet velocities and crossflow parameters ($V_{air}$, Pressure) attaining quicker stabilization, producing smaller, more uniform droplets.
- At low Weber numbers, with increasing momentum flux ratio (q), the droplet breakup is dominated by inertial and surface tension forces, resulting in minimal variation in $D_{0.999}$. At high Weber numbers, stronger aerodynamic forces enhance atomization and increase $D_{0.999}$ sensitivity to momentum flux ratio (q), with significant reductions as q rises.
- Increased aerodynamic forces at higher crossflow pressures and velocities accelerate the breakup process, shortening the breakup location and leading to improved droplet uniformity further downstream.
- Droplet number density analysis reveals that increased jet velocities and crossflow parameters improve atomization but reduce droplet residence time and alter distribution. Elevated crossflow pressures intensify droplet deformation, delaying the Eulerian-to-Lagrangian conversion and causing spray plume confinement. These factors, plume confinement, delayed particle conversion, and shorter residence time, collectively lead to the observed reduction in droplet density at elevated crossflow pressures.
- The Log-Normal distribution outperforms the Normal distribution in fitting droplet size data, as it effectively captures the skewness and tail behavior characteristic of atomization processes. In contrast, the Normal distribution provides a symmetric fit that is less accurate, particularly for asymmetric data. However, the Log-Normal distribution does not replicate the sharp peaks observed in the data, which are attributed to simulation-specific conversion criteria.
- The Rosin-Ramler distribution effectively represents droplet size trends in Cases 6 and 11. The minimal differences between its standard and modified versions are due to the limited skewness of the datasets and the standard model's inherent flexibility. While the Rosin-Ramler distribution effectively characterizes intermediate and larger droplet sizes, it slightly underestimates the smaller droplets.

The synergistic effects of momentum flux ratio, crossflow conditions, and Weber number provide a robust framework for optimizing atomization. This study enhances the understanding of LJICF systems through validated models and parametric analyses, offering valuable insights into tailoring spray atomization performance. Future work can extend this framework to investigate the impact of fluid viscosity, nozzle geometry, and evaporation effects on droplet dynamics, distribution, and atomization efficiency.

**Acknowledgments**

Financial support for this research is provided through the Science and Engineering Research Board, India, and the Department of Science and Technology (DST) under the National Supercomputing Mission (NSM), India. We acknowledge the National Supercomputing Mission (NSM) for providing computing resources of 'PARAM Sanganak' at IIT Kanpur, which is implemented by C-DAC and supported by the Ministry of Electronics and Information Technology (MeitY) and Department of Science and Technology (DST), Government of India. Also, we would like to thank the computer center (www.iitk.ac.in/cc ) at IIT Kanpur for providing the resources to carry out this work.

## AUTHOR DECLARATIONS

### CONFLICT OF INTEREST

The authors have no conflicts to disclose.

### DATA AVAILABILITY

The data that support the findings of this study are available from the corresponding author upon reasonable request.

### APPENDIX A: MEAN DIAMETERS AND DISPERSION PARAMETERS

Representative diameters provide a concise means of characterizing droplet size distributions in sprays. Metrics such as the mass median diameter ($D_{0.5}$) and upper-percentile diameters ($D_{0.9}$, $D_{0.999}$), are widely used in applications like fuel injection, spray cooling, and agriculture, where droplet size influences performance. Typically, a mean diameter and a dispersion measure are required to describe a distribution effectively. Several key representative diameters used to characterize droplet distributions based on volume are listed below:

TABLE II. Description of various diameter symbols used in the present study

| Representative diameters | Description |
|---|---|
| $D_{0.1}$ | Diameter at which 10% of the total liquid volume is in smaller droplets. |
| $D_{0.5}$ | Mass Median Diameter (MMD): 50% of the total liquid volume is in smaller droplets. |
| $D_{0.632}$ | Characteristic Mean Diameter: Diameter corresponding to 63.2% of the total liquid volume, related to cumulative distributions. |
| $D_{0.9}$ | Diameter at which 90% of the total liquid volume is in smaller droplets. |
| $D_{0.999}$ | Diameter at which 99.9% of the total liquid volume is in smaller droplets, indicating the distribution's tail, the largest droplet size present in the spray; an indicator of incomplete atomization or droplet coalescence downstream. |

| $D_{peak}$ | Diameter at the peak of the frequency distribution curve, indicating the most common droplet size. |
|---|---|

Additional key diameters include the Surface Mean Diameter ($D_{20}$), which emphasizes larger droplets and is crucial for surface-area-driven processes like evaporation and combustion, and the Volume Mean Diameter ($D_{30}$), important for volume conservation in applications such as fuel injection and flow control.

Droplet size dispersion parameters offer valuable insights into the spread and variability of droplet sizes within a spray, which is crucial for optimizing atomization performance across applications. The *Standard Deviation (STD)*, derived as the square root of variance, provides a basic measure of droplet size spread, indicating how widely sizes vary within the distribution.[28] The *Droplet Uniformity Index (DUI)***,** proposed by Tate[58], quantifies the degree of uniformity in droplet sizes, with higher values indicating a more consistent distribution. The *Relative Span Factor* ($\Delta_s$) measures the droplet size variability, with larger values indicating a broader dispersion, and is particularly useful for assessing sprays with diverse size distributions.[58] Similarly, the *Dispersion Boundary Factor* ($\Delta_B$) quantifies the range between median and upper-bound droplet sizes, providing insight into distribution width and atomization efficiency. Additionally, the *MMD/SMD* ratio (Mass Median Diameter to Sauter Mean Diameter) is a commonly used metric for dispersion, as it highlights the balance between larger and smaller droplets, critical in applications such as combustion and coating where precise control of droplet size is essential.

$$STD = \frac{\sum_i (D_i - D_{10})^2}{N} \tag{A1}$$

$$DUI = \frac{\sum_i V_i (D_{0.5} - D_i)}{D_{0.5}} \tag{A2}$$

$$\Delta_s = \frac{D_{0.9} - D_{0.1}}{D_{0.5}} \tag{A3}$$

$$\Delta_B = \frac{D_{0.999} - D_{0.5}}{D_{0.5}} \tag{A4}$$

In jet-in-crossflow systems, primary and secondary breakup shape the size distribution of the spray. Key factors include the Weber number (We) and momentum flux ratio (q), which influence jet penetration, droplet formation, and the size of the resulting spray plume. Diameters like $D_{21}$ and $D_{32}$ capture the effects of inertia, surface tension, and drag, with $D_{32}$ being particularly important as it indicates how efficiently droplets can evaporate during spray injection.

**APPENDIX B: DISTRIBUTION FUNCTIONS**

Standard distribution functions such as Normal**,** Log-Normal**,** and Log-Hyperbolic models facilitate comparisons between theoretical predictions and experimental data by smoothing statistical fluctuations. The normal distribution describes the random occurrence of a given drop size and applies to unbiased, random processes. It is expressed as a number distribution function:

$$f(D) = \frac{1}{\sigma\sqrt{2\pi}} e^{\left[-\frac{(D-\mu)^2}{2\sigma^2}\right]} \tag{B1}$$

where μ is the mean diameter, σ is the standard deviation, and $\sigma^2$ is the variance. The standard normal curve has an area of 1 under the curve, equally split on either side of the y-axis. The integral of this curve gives the cumulative number distribution function. The log-normal distribution is often used to model asymmetric distributions in sprays. Jet-in-crossflow atomization may produce such distributions due to non-uniform breakup and coalescence effects downstream, and is expressed as

$$f(D) = \frac{1}{\sqrt{2\pi}\sigma_{LN} D} e^{-\frac{[\ln(D)-\mu_{LN}]^2}{2\sigma_{LN}^2}} \tag{B2}$$

$$D_{10} = e^{\mu_{LN}+\sigma^2{}_{LN}/2} \tag{B3}$$

$$STD^2 = e^{2\mu_{LN}+\sigma^2{}_{LN}}(e^{\sigma^2{}_{LN}} - 1) \tag{B4}$$

Empirical models, including the Rosin-Rammler Distribution[48], Nukiyama-Tanasawa distribution[37], and Upper-Limit Function[36], are commonly used to model drop size distributions in atomized sprays. The Rosin-Rammler distribution, which is widely used for single-peaked distributions, provides cumulative probabilities based on a characteristic droplet diameter and spread parameter, but is less effective for multipeaked distributions. The Rosin-Rammler Distribution assumes a cumulative distribution function (CDF):

$$1 - Q = \exp - (D/X)^q \tag{B5}$$

where $Q$ is the fraction of the total volume contained in drops of diameter less than D, $X$ is the characteristic droplet diameter, and q is a constant that represents the spread of the distribution. This model is relevant for jet-in-crossflow systems where secondary breakup leads to a predictable range of droplet sizes. A significant drawback of this distribution function is the inability to deal with multiple-peaked distributions. However, it is quite useful for typical single-peaked results.